\documentclass[11pt]{article}
\usepackage[margin=1in]{geometry}

\usepackage{algorithm2e}
\usepackage{graphicx}
\usepackage{amsfonts}
\usepackage{dsfont}
\usepackage{amsmath,amsthm, amssymb}
\usepackage{bm}
\usepackage{braket}
\usepackage{hyperref}
\usepackage{mathtools}
\usepackage{xspace}
\usepackage[lambda,landau,operators,probability,sets,logic,complexity,asymptotics]{cryptocode}
\usepackage{dsfont}
\usepackage{tikz}
\usepackage{tkz-euclide}
\usepackage[capitalise]{cleveref}
\usepackage{subcaption}
\usepackage{comment}

\usepackage{float}
\usepackage{pgfplots}
\pgfplotsset{compat=1.18}
\usepackage{xargs}                      
\usepackage[colorinlistoftodos,prependcaption,textsize=tiny]{todonotes}

\usepackage[normalem]{ulem}
\useunder{\uline}{\ul}{}

\makeatletter
\newtheorem*{rep@theorem}{\rep@title}
\newcommand{\newreptheorem}[2]{%
	\newenvironment{rep#1}[1]{%
		\def\rep@title{#2 \ref{##1} (restated)}%
		\begin{rep@theorem}}%
		{\end{rep@theorem}}}
\makeatother

\newcommand{\petros}[1]{#1}
\newcommand{\milos}[1]{#1}
\newcommand{\mycomment}[1]{}

\newcommand\scalemath[2]{\scalebox{#1}{\mbox{\ensuremath{\displaystyle #2}}}}
\input{Qcircuit}	

\newtheorem{thm}{Theorem}
\newtheorem*{thm*}{Theorem}

\newtheorem*{lem*}{Lemma}

\newtheorem{prop}[thm]{Proposition}

\newreptheorem{thm}{Theorem}
\newreptheorem{lem}{Lemma}
\newreptheorem{prop}{Proposition}
\newreptheorem{cor}{Corollary}

\DeclareMathOperator{\probability}{\mathds{P}}
\DeclareMathOperator{\hamiltonian}{\mathcal{H}}
\DeclareMathOperator{\lattice}{\mathcal{L}}

\renewcommand{\vec}[1]{\ensuremath{\boldsymbol{#1}}\xspace}
\newcommand{\mat}[1]{\ensuremath{\boldsymbol{#1}}\xspace}

\DeclareMathOperator{\mR}{\mathbb{R}}
\DeclareMathOperator{\mZ}{\mathbb{Z}}
\DeclareMathOperator{\cL}{\mathcal{L}}

\DeclareMathOperator{\basis}{\mathbf{B}}

\makeatletter
\newcommand{\vast}{\bBigg@{3}}
\newcommand{\Vast}{\bBigg@{5}}
\makeatother

\title{Heuristic Time Complexity of NISQ Shortest-Vector-Problem Solvers}

\author{
    Milo\v{s} Prokop\thanks{\texttt{m.prokop@sms.ed.ac.uk}}, 
    Petros Wallden\thanks{\texttt{petros.wallden@ed.ac.uk}} \\
    School of Informatics, University of Edinburgh
}

\date{}

\excludecomment{draft}

\begin{document}
	
\maketitle

\begin{abstract}
	Finding the shortest non-zero vector on a lattice is a fundamental problem in lattices that is believed to be hard both for classical and quantum computers. Two of the three standardised by NIST post-quantum cryptosystems rely on its hardness. Research on theoretical and practical performance of quantum algorithms to solve the Shortest Vector Problem (SVP) is crucial to establish confidence in those new standards. Exploring the capabilities that Variational Quantum Algorithms (VQA) that can run on Noisy Intermediate Scale Quantum (NISQ) devices have in solving SVP has been an active area of research. The qubit-requirement for doing so has been analysed and demonstrated that it is plausible to encode SVP on the ground state of a Hamiltonian in an efficient way. Due to the heuristic nature of VQAs no analysis of the performance and specifically of the time complexity of those approaches for scales beyond the non-interesting classically simulatable sizes has been performed. In this paper, motivated by Boulebnane and Montanaro \cite{Boulebnane:2022ace} work on the k-SAT problem, we propose to use angle pretraining of the Quantum Approximate Optimisation Algorithm (QAOA) for SVP and we demonstrate that QAOA with fixed angles performs well on much larger instances than those used in training. Avoiding the limitations that arise due to the use of optimiser, we are able to extrapolate the observed performance and we observe the probability of success scaling as $2^{-0.695n}$ with $n$ being dimensionality of the search space for a depth $p=3$ pre-trained QAOA algorithm. By repeating the algorithm, we get \petros{heuristic} time complexity $O(2^{0.695n})$ to solve the problem with high probability, a bit worse than the fault-tolerant Grover approach of $O(2^{0.5n})$. On the other hand, both the number of qubits, and most importantly the depth of each quantum computation, are considerably better in our approach – Grover requires exponential depth, while each run of constant $p$ fixed-angles QAOA requires polynomial depth. Further, we analyse the performance of QAOA in solving approximate-SVP, a version of the problem that one seeks a multiplicative approximation to the shortest vector. As a side result, we propose a novel method to avoid the zero vector solution to SVP without introducing more logical qubits. This improves upon the previous works as it results in more space efficient encoding of SVP on NISQ architectures without ignoring the zero vector problem.
\end{abstract}	
    
\section{Introduction}
The Shortest Vector Problem (SVP) is a fundamental mathematical problem on Euclidean lattices, that is conjectured to be hard both for classical and quantum computers. Because of this, cryptosystems that base their security on the hardness of SVP, have been developed and appear as leading candidates for the new post-quantum cryptographic standards \cite{bos2018crystals,ducas2018crystals}. 
The task in SVP is the following: given a basis (defining) a lattice to find the shortest non-zero vector in the lattice. The concrete hardness of the different instances of the problem depend on the dimension, the structure and whether an exact or approximate solution is required. To use a problem as basis for cryptographic solutions, one needs on the one hand to have confidence that there is no classical or quantum algorithm that can efficiently (in polynomial time) solve the problem. There is strong evidence for SVP, for example its exact version has been proven NP-hard under randomised reductions in the $\ell_2$ norm  \cite{ajtai1998, micciancio2001}. On the other hand, to determine the concrete security of a cryptosystem and to set the key-sizes for standardised implementations, one needs to know the best classical and quantum algorithms, even if their scaling is exponential, and at the same time have confidence that whatever improvements or new records for solving the problem, are not going to drastically change the evaluation -- ensuring that key sizes chosen will provide security even after such developments.
The best classical algorithms can provably solve SVP in $2^{2.465n+o(n)}$ time with $2^{1.233n+o(n)}$ space \cite{fasterSieving2} using lattice sieving. Under heuristic assumptions, a long line of works \cite{BGJ13,laarhoven2015search,La15,sievingRecord} decreased the time complexity to $2^{0.292n+o(n)}$ time with $(4/3)^{n/2+o(n)}$ space. The fastest known quantum fault-tolerant (FT) SVP solver runs in time \(2^{0.257\,n+o(n)}\), uses QRAM of maximum size \(2^{0.0767\, n+o(n)}\), a quantum memory of size \(2^{0.0495\,n+o(n)}\) and a classical memory of size \( 2^{0.2075\,n+o(n)}\) \cite{chailloux2021}. However, unless further breakthroughs are discovered, the best quantum sieving algorithms would still, under very optimistic assumptions, take $10^{31}$ years to solve SVP on 400 dimensional lattice, roughly the minimal dimension required for post-quantum cryptographic protocols standardised by National Institute for Standards and Technology \cite{doriguello2024}.
The quantum enumeration approaches are alternative to sieving approaches. They combine exhaustive search with projections into lower dimensional lattices, e.g. \cite{AC:AonNguShe18}. These algorithms run in worse asymptotic time \(n^{n/(4e) + o(n)}\), but have the advantage that they require \(\poly[n]\) memory in contrast to sieving and therefore would be a preferred choice if available memory does not allow the run of sieving algorithms. Furthermore, in many cryptographic applications the runtime of \(n^{n/16  + o(n)}\) seems plausible \cite{C:ABFKSW20}.

However, not much is known about the performance of SVP solvers on Noisy Intermediate Scale Quantum (NISQ) architectures, and of algorithms inspired by those architectures. 
Because of low qubit availability and high error rates, the use of full error-correcting techniques is not applicable in the near future for problems of large complexity. To circumvent the issue, specialised classes of quantum algorithms have been proposed called variational quantum algorithms (VQAs). The most notable examples are Variational Quantum Eigensolver (VQE) \cite{vqe1,vqe2} and Quantum Approximate Optimisation Algoritm (QAOA) \cite{farhi2014quantum}. The latter is especially suited for solving combinatorial optimisation problems, which naturally include the SVP. Potential vulnerabilities of post-quantum cryptographic protocols will be at first exploited by NISQ devices, therefore research on their capability in solving SVP is of a significant importance. Limited research on the topic exists \cite{notSoAdiabatic,JCLM21,vqaSvpMilos,zhu2022iterative,mizuno2024,Ura_2023}, but is focused on analysing the space-complexity, i.e. the number of qubits required, \milos{or SVP to Hamiltonian mapping strategies, }and not the time taken to solve the problem. The bottleneck of the studies is that VQAs are heuristic quantum algorithms, with the lack of theoretical results to extrapolate their real performance. In other words, their performance beyond the scales that can be experimentally evaluated on today's quantum hardware (or emulated on a classical hardware) is not well understood. These works have tested NISQ approaches to SVP on very small lattice sizes (up to 6 on real-world quantum hardware and up to 28 via classical emulation), and it is unclear how the observed performance extrapolates to much larger, cryptographically interesting, lattice dimensions. The major factor is that VQAs are vulnerable to \textit{barren plateaux} caused by exponentially ``vanishing'' gradients of the optimised cost function \cite{cerezo2021cost,McClean2018,Wang2021}. Although experiments performed with VQAs suggest interesting performances on small sized lattice instances \cite{JCLM21,vqaSvpMilos}, barren plateaux will eventually limit the effectiveness of these approaches for larger, cryptographically more interesting dimensions. At what scales and under what conditions they appear, is unclear and an open research question. This significantly complicates any extrapolation of the observed VQA based SVP solvers.

This paper introduces a so-called Fixed-angle QAOA and Fixed-angle CM-QAOA approaches that use pre-trained ansatz parameters from small instances to solve much larger instances without a need to run the full, optimised version of QAOA (or CM-QAOA) algorithm. Without the need to perform optimisation (because of the pre-trained angles), the algorithms are no longer vulnerable to barren plateaux, essentially removing the major obstacle to extrapolate and estimate their performance on large instances. This is obviously an underestimation of performance of QAOA-based SVP solvers on large instances (as it constraints the algorithm by removing the optimisation part), but as we show in \Cref{sec:exactSVP} these underestimations still provide interesting scalings that upper bound the estimations of the NISQ SVP enumeration approaches. In this paper, for first time, we address this issue by giving some reliable bounds on the time-complexity of NISQ SVP solvers, that apply to larger dimensions. For this we use Fixed-angle QAOA (and Fixed-angle CM-QAOA introduced later) that leverage the need to use an optimiser to solve each new unseen instance of SVP. This method was recently introduced by Boulebnane and Montanaro \cite{Boulebnane:2022ace}, who used pre-trained QAOA angles for the $k$-SAT problem, and demonstrated that, for $k=8$, the pre-trained QAOA of depth 14, matches the performance of the best known classical $k$-SAT solver and significantly outperforms the naive use of Grover's algorithm. Here, we follow similar approach by finding \textit{good} QAOA angles that solve the small training instances with high probability,  in comparison with random guess. We show that performance of Fixed-angle QAOA and Fixed-angle CM-QAOA can be reliably extrapolated into higher dimensions with these angles. Although in small lattice instances, the performance of Fixed-angle (CM-)QAOA is strictly worse than those of optimised, standard version, (CM-)QAOA, we show that the observed asymptotic scaling of the performance is interesting because (i) it provides upper bound for the performance of (CM-)QAOA solvers and (ii) we show that for very low constant algorithm depths $p\leq3$ with $O(n\log n)$ qubits, the observed performances approach the performance of Grover's algorithm, which is a FT algorithm with much larger qubit and circuit depth requirements. The interesting aspect of this comparison is the finding that Grover's algorithm approach to SVP requires $\Theta\left(\sqrt{2^{(3/2)n\log n}}n^2\log^2 n\right)$ deep quantum circuits and $\Theta(n^2\log n)$ qubits \cite{prokop2024groversoracleshortestvector} while experiments show that Fixed-angle (CM-)QAOA solves the problem with $poly(n)$ depth of quantum circuit and $O(n\log n)$ qubits to achieve a probability of success $O(2^{-\alpha n})$ with $\alpha$ potentially approaching $1/2$ for larger depths $p$, mimicking Grover\footnote{Probability $O(2^{-\alpha n})$ of finding a solution means that by repeating $O(\alpha n)$ times we can solve SVP with high probability. These repetitions can happen sequentially or in parallel. Note that a \textit{single} run of our QAOA has polynomial circuit depth and the repetitions do not even need to be performed at the same quantum computer, while Grover-based approaches require exponential depth quantum circuit, since the iterations need to happen sequentially on the same quantum circuit.}. 

In this paper we discuss pretraining method for Fixed-angle QAOA and Fixed-angle CM-QAOA, evaluate how well the trained angles performed on larger unseen instances and estimate the scaling of the algorithms beyond lattice sizes that can be emulated classically. \milos{We emphasise that our time complexity results are conjectured based on extrapolation from experimental data on lattice dimensions up to 22, with training performed on dimensions 4 to 10. Although we provide support for the scalability of our approach, these complexity bounds should be understood as heuristic estimates rather than rigorous analytical results.} Finally, we compare and discuss the performance of Fixed-angle QAOA and Fixed-angle CM-QAOA both on Exact-SVP and Approximate-SVP to present more detailed estimates of the performance of Fixed-angle (CM-)QAOA that can be better compared to requirements posed by cryptographic protocol proposals.

As for a second major contribution of the paper, we propose a new solution to avoid convergence towards a trivial, non-interesting, zero vector solution, which gets naturally encoded as the global optimum of the problem to be solved by a method to translate SVP into Ising spin Hamiltonian form. The problem has been ignored in \cite{notSoAdiabatic,JCLM21} with the reasoning that the actual SVP solution may still be returned with a high probability even though it is not the globally optimal solution given their problem encodings. \cite{vqaSvpMilos} proposed a different method by modification of the cost function that provably results in a problem formulation whose ground state is the solution to SVP. However, such a solution requires to introduce extra $n-2$ logical qubits which can be too costly for the context of NISQ devices. This paper proposes a different method to avoid the zero vector solution by modifying QAOA resulting in an algorithm that we call a CM-QAOA (controlled mixer QAOA). It is an instance of a generalised QAOA called Quantum Alternating Operator Ansatz \cite{quantumAlternatingOperatorAnsatz} and it is used to constraint any leakage of the amplitude of the state of the system to the zero vector solution. In contrast to previous approaches, CM-QAOA does not increase qubit count and provably leads to an algorithm that converges towards the shortest non-zero lattice vector as the depth $p$ increases. However, the depth \(p\) required to achieve a reliable solution in practice—particularly how it scales with the lattice dimension—is currently unknown and remains as an open question subject to further experimentation.

\subsection{Contributions}\label{sec:contributions}
\begin{itemize}
    \item We experimentally estimate and extrapolate upper bound on time complexity of QAOA based SVP solver. We use the fact that QAOA performs at least as good as Fixed-angle QAOA and consequently an extrapolation on Fixed-angle QAOA-based SVP solver performance provides an upper bound on time complexity for solving SVP via QAOA. Specifically, we show that, for depth $p=3$, QAOA and CM-QAOA have \milos{conjectured} time complexities of solving SVP upper bounded by \(2^{0.695n+o(n)}\) and \(2^{0.895n+o(n)}\) respectively.

    \item We propose a novel way to exclude the zero vector solution from the search space of the SVP variational algorithm solver based on QAOA. Our novel method, called CM-QAOA (controlled mixer QAOA) is a usage of Quantum Alternating Operator Ansatz with specific mixing unitaries that do not increase an overlap of the quantum system's statevector with the zero vector solution. CM-QAOA is introduced in \Cref{sec:constrainingWithFixQAOA}.
	\item We introduce pre-training of QAOA angles for SVP instances\footnote{Whenever we mention QAOA for SVP we mean running QAOA with Hamiltonian $\mathcal{H}$ obtained by mapping the SVP problem to $\mathcal{H}$ without introducing a penalisation for the zero-vector solution as outlined in \cite{vqaSvpMilos}} and show that there exist sets of angles that universally provide advantage across classes of SVP instances. These angles can either be used as initial angles for SVP solvers based on QAOA or can be used in non-optimising version of QAOA, in Fixed-angle QAOA (presented in \cref{sec:fixedAngleQAOA}). We experimentally estimated the advantage of using Fixed-angle QAOA and Fixed-angle CM-QAOA and the results are presented in \cref{sec:expResults}.
	\item We design a method for generating SVP-instances of small sizes that are capable of being converted to QUBO formulations and run in classical quantum emulations of variational quantum algorithms, and yet which mimic hardness of SVP for the variational quantum algorithms in such small instances.
    \item In \cref{sec:approxSVP} we evaluate the performance of Fixed-angle QAOA and Fixed-angle CM-QAOA on approximate SVP problem $\gamma$-SVP with $\gamma=\text{poly}(n)$. This evaluation is important in the context of real-world cryptographic instances as many of the current post-quantum proposals rely on conjectured hardness of $\gamma$-SVP with polynomial factor $\gamma$.
\end{itemize}

\noindent \textbf{Outline}. The rest of the paper is structured in the following way. \Cref{sec:RW} lists the related works and \Cref{sec:prelim} presents the preliminaries. \Cref{sec:paramPretraining} outlines our method, the Fixed-angle QAOA, for pre-training of QAOA angles for SVP problem. \Cref{sec:constrainingWithFixQAOA} proposes a novel way, the CM-QAOA, to exclude the zero vector solution from the search space of the SVP variational algorithm solver without a need to utilise additional logical qubits. \Cref{sec:exactSVP} describes experimental approach to test performance of Fixed-angle QAOA and Fixed-angle CM-QAOA algorithms on SVP and present the results that upper bound the expected performance of QAOA based SVP solvers. \Cref{sec:approxSVP} evaluates performance of Fixed-angle QAOA and Fixed-angle CM-QAOA on approximate \(\gamma\)-SVP with \(\gamma=\poly(n)\). We conclude in \Cref{sec:cmQaoaConclusions}.

\subsection{Related Work}\label{sec:RW}
\subsubsection{Analysing the performance of QAOA}\label{sec:RWQAOAPerf}
The time complexity of QAOA is difficult to estimate and is an open research question. The reasons include occurrence of barren plateaus \cite{McClean2018,Wang2021} that cause exponential difficulty of optimising the QAOA induced by noise present in an underlying quantum architecture.  \cite{Shaydulin2019MultistartMF} argues that another difficulty in finding the optimal QAOA angles is due to many local optima. Therefore the authors suggest the use of multistart methods when instances with different initial angles are being optimised. Moreover, the authors demonstrated that optimal angles found for a particular problem can be re-used as an initial point to similar problems. This finding improves both the quality of the solution and reduces the time complexity. It also supports the use of Fixed-angle QAOA presented in \Cref{sec:fixedAngleQAOA} by showing that similar problems may have ``similar'' optimal angles. Because of the difficulty in time complexity analysis of QAOA, most of the analysis found in literature is focused on specific applications of QAOA that also require some modifications to the algorithm resulting in many different variations of QAOA that are easier to analyze, for instance Reinforcement Learning Assisted Recursive QAOA \cite{Patel2022ReinforcementLA}, Adapting QAOA \cite{PhysRevResearch.4.033029} or Multi-angle QAOA \cite{multiAngleQAOA}. Each work has a specific approach to evaluate time complexity of the algorithm it proposes. \cite{CroocksMaxcut} explores the performance of QAOA by benchmarking it against classical algorithms, discussing the trade-off between depth of QAOA, quality of the solution and its time complexity on Max-Cut problem. The author found that with modest circuit depth QAOA is able to outperform the classical polynomial time Goemans-Williamson algorithm and that performance with fixed QAOA depth was insensitive to size of problems. \cite{Jiang:2017ifw} designs a QAOA-based unstructured search algorithm and by introducing a technique based on spin-coherent states it proves its quadratic speedup that matches the speedup of the Grover's algorithm.

\subsubsection{Penalising the trivial zero vector SVP solution}\label{sec:RWPenalisationZeroVect}
For the purpose of solving the SVP problem in the NISQ era, the suitable VQAs capable of performing discrete optimisation with real-valued cost are Variational Quantum Eigensolver (VQE) \cite{vqe1,vqe2} and QAOA \cite{farhi2014quantum}. Solving SVP via VQAs requires to first convert it into a Hamiltonian operator, whose ground state is the solution to SVP. However, this is not trivial to achieve as the shortest vector of any lattice is the trivial zero vector and it needs to be included. In other words, SVP is asking for a shortest non-zero lattice vector, which is one of the vectors of the lattice with the second smallest length. \cite{notSoAdiabatic,JCLM21} proposed to ignore the problem for very small lattice instance sizes and use the fact that quantum algorithms may return sub-optimal solutions, which can be the solution to SVP, with a non-negligible probability. Such a solution is however not expected to work well as the lattice sizes scale. \cite{vqaSvpMilos} suggested that VQE has advantage that the cost may be modified classically to penalise the trivial zero-vector solution of every SVP formulation. However, VQE has no theoretical guarantee to converge to the solution. QAOA on the other hand mimics adiabatic quantum evolution and in the limit of its depth approaching infinity, it is guaranteed to return the optimal solution. When applied to SVP it is not possible to penalise the zero vector classically if the theoretical guarantee to find the solution to SVP is not to be given up. Instead, one needs to modify the problem Hamiltonian as was suggested in \cite{vqaSvpMilos} which increases qubit requirements by $n-2$ for an $n$ dimensional lattice. The cost to encode constraints in the problem Hamiltonians motivated Hadfield et al. \cite{quantumAlternatingOperatorAnsatz} to generalise QAOA by consideration of a more general class of mixing unitaries instead of applying $e^{-i\beta_iH_M}$ with $H_M=\sum_{i}^{n}X_i$ being in each of the QAOA layers. They called the generalised algorithm Quantum Alternating Operator Ansatz with the same acronym QAOA and argue that it allows for a more efficient implementation of hard constraints. It motivated construction of a mixing unitary that we propose in this paper that helps QAOA to avoid the zero vector solution without any further qubit requirements, although with a relatively small cost of increased gate count and circuit depth.

\subsubsection{Pretraining of QAOA}\label{sec:fixedAngleQAOA}
Because the optimization subroutine that finds the optimal parameters of QAOA can be computationally expensive \cite{Shaydulin2019MultistartMF,9951269,Medvidovic2021}, methods to leverage the QAOA parameter search are being investigated. \cite{EvaluatingQAOA,ParamConcentrations,Brando2018ForFC,9605328} suggest that the optimal angles can be reused for other instances of the same problem class. Machine learning models can be utilised for predicting optimal parameters for certain problem classes \cite{reinforcementLearningBased,learningToOptimize,10.5555/3408352.3408509}. Some problem classes even allow analytical calculation of the optimal angles, e.g. triangle-free regular graphs at QAOA depth $p=1$ \cite{PhysRevA.97.022304}, Sherrington-Kirkpatrick model  \cite{Farhi2022quantumapproximate} or MAX-CUT instances on Erdos-Renyi random graphs with number of vertices $n\rightarrow\infty$ \cite{Boulebnane:2021sda}. Alternatively, the optimal parameters can be experimentally estimated as in \cite{Boulebnane:2022ace}, where authors estimated optimal QAOA angles for k-SAT problem. Because using pretrained angles results in an algorithm whose performance matches performance of QAOA at best, the analysis with fixed QAOA angles provides lower bound on QAOA performance. These lower bounds were sufficient to prove analytic bounds on averaged performance of QAOA over random boolean formulae at the satisfiability threshold as the number of variables $n\rightarrow\infty$. The authors found that QAOA depth $p=14$ is sufficient to match the performance of the state-of-the-art WalkSATlm solver. It is expected that for $p>14$ the performance of fixed-angle QAOA would overperform WalkSATlm although the amount of actual advantage needs further study.

In parallel to our work, Priestley and Wallden \cite{priestley2025} have proposed a pretrained QAOA approach for solving the Closest Vector Problem (CVP), a related lattice problem to SVP. While their primary focus is on CVP with structured ``prime'' lattices used for integer factorisation, our work concentrates solely on SVP (and its approximate version), which is more directly relevant to post-quantum cryptographic schemes. The authors use pretraining approach based on selecting angle sets that optimise scaling performance on small random instances and validating them across larger instances. Notably, the authors perform refinement within a local neighbourhood around an approximate solution obtained via Babai’s nearest plane algorithm \cite{Babai1986OnLL}, training QAOA angles to increase the probability of finding a better solution in that restricted region. By contrast, we pretrain angles to directly maximise the overlap with the shortest non-zero vector in global SVP instance dataset, providing a broader view of algorithm performance beyond local refinement and potentially enabling better generalisation. Moreover, we also extend our analysis of pretrained QAOA to approximate variant of SVP, the $\gamma$-SVP and introduce a constrained-mixer QAOA variant (CM-QAOA) that eliminates the zero-vector solution without increasing qubit count, which is not relevant for the CVP.

\section{Preliminaries}\label{sec:prelim}
\subsection{Lattices and post-quantum cryptography}\label{sec:prelimLattices}
\emph{Lattice} $\cL$ is the set $\cL(\vec{b}_{1},\dots,\vec{b}_{n}):=\left\{ \sum_{i=1}^{n}x_{i} \vec{b}_{i} ~:~ x_{i}\in\mZ\right\} $ of all integer combinations of $n$ linearly independent vectors $\vec{b}_{1},\dots,\vec{b}_{n} \in \mR^{d}$. $\vec{b}_1,\dots,\vec{b}_n$ are called a \emph{basis} of $\cL$ and there are infinitely many possible bases for a particular lattice. A lattice basis is usually denoted on a form of a matrix $\mat{B}$ whose rows form a basis of $\cL$. All the bases of a lattice contain the same number of elements $n$ that is called the \emph{dimension} or \emph{rank} of $\cL$.  The set of all vectors $\vec{x}\in\operatorname{span}(\cL)$ such that $\vec{x}\cdot\vec{y}\in\mathbb{Z}$ is an integer for all $\vec{y}\in \cL$ is called the \textit{dual lattice} $\widehat{\cL}$ of a lattice $\cL$. A Gram matrix is defined as $\mat{G}:=\mat{B}\mat{B}^T$ and it allows to list the squared lengths of vectors in $\cL$ by enumerating over all row-vectors $\vec{x}\in\mathbb{Z}^n$: if $y=\vec{x}\cdot \mat{B}$ then $\|\vec{y}\|^2=\vec{x} \cdot \mat{B}\cdot \mat{B}^T\cdot \vec{x}^T=\vec{x}\cdot \mat{G} \cdot \vec{x}^T$. \textit{Successive minima} of $\cL$ list the lengths of the vectors of $\cL$: $\lambda_0=0$ by convention, $\lambda_1$ denotes the length of (any of) the shortest non-zero vector(s) of $\cL$, $\lambda_2$ denotes the length of (any of) the second shortest non-zero vector(s) of $\cL$ and so on. The \textit{Shortest Vector Problem} (SVP) asks for a solution to $	\lambda_1 \coloneqq \min_{\vec{y}\in\mathcal{L}(\mat{B})\setminus\{\vec{0}\}}\|\vec{y}\|$ and the approximate version $\gamma$-SVP asks for a vector of a length $\leq \gamma \lambda_1$. Proven or conjectured NP-hardness of particular SVP variants has been fundamental to construction of many post-quantum cryptography proposals. Out of four finalists chosen from six-year competition run by US National Institute of Standards and Technology, the hardness of three of them is based on lattice problems \cite{nist2022quantum}: CRYSTALS-Kyber \cite{bos2018crystals} for key encapsulation mechanism and CRYSTALS-Dilithium \cite{ducas2018crystals} and FALCON \cite{fouque2018falcon} for signatures.

\subsection{Lattice basis reduction algorithms}

Lattice basis reduction algorithms reduce a basis $\mathbf{B}$ into a basis $\mathbf{B}'$ improving on shortness of the basis vectors and/or their mutual orthogonality and are also employed to solve the SVP. They are generally categorised as enumeration and sieving methods. The enumeration methods list all vectors of $\cL$. and typically pruning strategies are used to limit the search space and improve their running time. Classically, the cost of the enumeration algorithms is $2^{O(n \log n)}$\footnote{Unless a base of a logarithm is specified, $log(\cdot)$ is a logarithm with base 2.}\cite{Kannan1987,kannanImproved,SchnorrEuchner1994} on an $n$-dimensional lattice and requires a polynomial space. The sieving methods work by progressive build-up of large sets of lattice vectors and performing mutual operations to obtain lattice vectors of shorter lengths. They run asymptotically faster, the fastest one taking $2^{0.292n + o(n)}$ time \cite{laarhoven2015sieving}, though the sieving algorithms require an exponential space. Typically enumeration outperforms sieving for small problem sizes until a crossover point around at around dimension $70$ \cite{albrecht2019general}. That is to say, despite better asymptotic performance, enumeration is still a better choice for some practical choices of $n$. To asses the quality of lattice basis reduction algorithms, two metrics on the reduced basis $\mathbf{B}'=(\vec{b}'_1,\dots,\vec{b}'_n)^T$, with $\|\vec{b}_1\|,\|\vec{b}_2\|,\dots$ in non-descending order, are usually being considered: 1. \textit{Approximation factor} that relates the length of the shortest returned non-zero vector with the length of the actual shortest non-zero vector of $\mathcal{L}$, i.e. $|\mathbf{b}'_1|/\lambda_1(\mathcal{L})$ 2. \textit{Hermite factor} that relates the length of the shortest returned non-zero vector with an $n$-th root of the volume of $\mathcal{L}$, i.e. $|\mathbf{b}'_1|/\text{vol}(\mathcal{L})^{1/n}$.

\subsection{Generalised QAOA algorithm}
In 2014 Farhi et al. \cite{farhi2014quantum} proposed Quantum Approximate Optimisation Algorithm (QAOA) which approximates optimisation via Quantum Adiabatic Algorithm \cite{adiabaticFarhi} in discrete steps.  It has been found to be well-suited for Noisy-Intermediate Scale Quantum (NISQ) architectures due to shallow quantum circuit depths and leverage on quantum capabilities by delegating the computation between classical and quantum devices. Its goal is to find a ground state (i.e. the lowest energy eigenstate) $\ket{\psi_\text{GS}}$ of a cost Hamiltonian operator $\hamiltonian_C=\sum_x f(x)\ket{x}\bra{x}$ whose eigenstates correspond to a search space defined by an image of a classical function $f: \mathbb{Z}^n\rightarrow\mathbb{R}$ in an $n$-qubit system. It proceeds by defining a mixer Hamiltonian $\hamiltonian_M$ with a requirement that it does not commute with $\hamiltonian_C$. It defines a feasible domain of solutions and $\hamiltonian_M=\sum_{i=1}^nX_i$ which allows a search over all possible $n$-bit bitstrings has been mostly adapted. \cite{PhysRevLett.124.090504,Golden_2023} show that such a choice may be non-optimal and there an improved performance of QAOA can be observed when a more optimal pair of $\hamiltonian_C$ and $\hamiltonian_M$ is chosen for both constrained and unconstrained optimisation problems. Farhi's QAOA \cite{farhi2014quantum} assume time evolution under a fixed local mixer Hamiltonian $\hamiltonian_M$. This is formally generalised in Quantum Alternating Operator Ansatz \cite{quantumAlternatingOperatorAnsatz} which considers a more general family of non-local mixing unitaries and argue that they allow for representation of more useful set of states which might be especially useful if the problem involves hard constraints. Let us denote by \(\mathcal{U}_M(\beta_k)\) a general mixing unitary parameterised by an angle \(\beta_k\). The optimisation is then being performed by preparing an initial state $\ket{s}=H^{\otimes n}\ket{0^{\otimes n}}$ and using a classical optimiser to find the optimal angles $\bm{\beta}=(\beta_1,\beta_2,\dots,\beta_p),\bm{\gamma}=(\gamma_1,\gamma_2,\dots,\gamma_p)$ that minimise the following quantity implemented as a quantum circuit and run on a quantum device:
$$
\min_{\bm{\beta},\bm{\gamma}}\mathcal{U}_M(\beta_p)\mathcal{U}_C(\gamma_p)\cdots \mathcal{U}_M(\beta_2)\mathcal{U}_C(\gamma_2)\mathcal{U}_M(\beta_1)\mathcal{U}_C(\gamma_1)\ket{s}
$$
with $\mathcal{U}_C(\gamma_j)\coloneqq e^{-i\gamma_j\hamiltonian_C}$. After the optimisation the resulting state is
$$
\ket{\psi_\text{QAOA}}=\prod_{j=1}^{p}\mathcal{U}_M(\beta_{p-j+1})e^{-i\gamma_{p-j+1}\hamiltonian_C}\ket{s}
$$ and as the QAOA depth $p\rightarrow\infty$ it is guaranteed by the Adiabatic Theorem that the overlap $\braket{\psi_\text{GS}|\psi_\text{QAOA}}\rightarrow 1$.

\subsection{Shortest Vector Problem (SVP) in the variational quantum algorithms framework}\label{sec:prelimSvpVqas}
A row-wise basis $B\in\mathbb{Z}^{d\times d}$ defines a corresponding lattice $\mathcal{L}(\mat{B})=\{\vec{x} \mat{B} |\vec{x}\text{ row vector}, \vec{x}\in\mathbb{Z}^d\}$. The shortest vector problem (SVP) asks to find a solution to $\lambda_1 \coloneqq \min_{\vec{y}\in\mathcal{L}(\mat{B})\setminus\{\vec{0}\}}\|\vec{y}\|$.
Let $\vec{x}$ be a $d$-dimensional row vector of coefficients and $\mat{G}=\mat{B} \mat{B}^T$ a Gram matrix of the basis. Then $\vec{y}:=\vec{x} \mat{B} \implies \|\vec{y}\|^2=\vec{x} \mat{B} \mat{B}^T \vec{x}^T=\vec{x}\mat{G} \vec{x}^T$ which allows us to reformulate SVP as a quadratic constrained integer optimisation problem \milos{\cite{vqaSvpMilos}:}
\begin{align}\label{eq:svpformulation2}
\lambda_1^2&=\min_{\vec{y}\in\mathcal{L}(\mat{B})\setminus\{0\}}\|\vec{y}\|^2\nonumber\\
&=\min_{\vec{x}\in\mathbb{Z}^n\setminus\{\vec{0}\}}\sum_{i=1}^nx_i^2 \mat{G}_{ii}+ 2 \sum_{1\leq i<j\leq n}x_i x_j  \mat{G}_{ij} 
\end{align}
In order to construct a corresponding Hamiltonian $\hamiltonian_C(\lattice)$ the Equation (\ref{eq:svpformulation2}) needs to be converted into quadratic unconstrained binary optimisation (QUBO) formulation. For this, bounds $|x_i|\leq \alpha_i$ need to be chosen to define a search space for the enumeration. Non-optimal search space size can either result in omitting the SVP solution from the search space or making the enumeration overly complex, increasing quantum resource requirements and decreasing the probability of finding the SVP solution. The appropriate scalings for the bounds $\bm{\alpha}$ are explored in \cite{vqaSvpMilos}. The authors also propose an algorithm to HKZ-reduce a lattice basis, one of the strongest notions of lattice reduction \cite{Kannan83}, using $\frac{3}{2}\log_2(d)-2.26d+O(\log_2 d)$ qubits. Alternatively, if the goal is evaluation of the performance of variational quantum algorithms on solving SVP one may choose sub-optimal bounds for the variables $x_i$ as was done e.g. in the experimentation part of \cite{vqaSvpMilos}. In this case it is the performance of finding the shortest non-zero vector in a search space that is evaluated, not the performance of finding the shortest non-zero lattice vector as it may be missing in the search space. Such a relaxation of the problem allows for better evaluation of variational quantum algorithms (by classical emulation or on a constrained quantum hardware) as otherwise, if $O(n\log n)$ qubits are used, only very small lattices can be classically emulated due to constraint on number of qubits. \milos{Other mappings of SVP to a Hamiltonian $\hamiltonian_C$ exist \cite{JCLM21,mizuno2024,dableheath2025} with different pros and cons depending on the underlying quantum architecture specifications, though they are not relevant for this paper as classical emulations are used instead and the aim is to minimise the number of logical qubits to fit into classical emulation limits.}

\subsection{Approximate Shortest Vector Problem}\label{sec:prelimApproxSVP}
The Approximate Shortest Vector Problem ($\gamma$-SVP) is a variation of SVP, where the goal is not to find the shortest non-zero lattice vector with the length $\lambda_1$, but rather to find a lattice vector with a length $\leq\gamma\lambda_1$. The version with $\gamma=1$ that asks for the exact non-zero shortest vector of a lattice is sometimes denoted as Exact-SVP to stress the nature of the problem. This problem is central to lattice-based cryptography, as it balances computational feasibility with security. $\gamma$-SVP is considered intractable for both classical and quantum computers for certain approximation factors, making it a key hardness assumption behind many post-quantum cryptographic schemes. $\gamma$-SVP is proven to be NP-Hard (under randomised reductions) for factors $\gamma\leq\sqrt{2}$ in $l_2$ norm (which is assumed throughout this work) \cite{hardInConstantApprox}. The complexity of $\gamma$-SVP within $\gamma=\text{poly}(n)$ is an open research area. Approximate SVP within polynomial factors underpins the security of several lattice-based cryptosystems like LWE \cite{regev2009lattices}, Ring-LWE \cite{lyubashevsky2010ideal} or Ajtai-Dwork cryptosystem \cite{ajtaiCryptosystemPolyApprox} whose assumption is that solving $\gamma$-SVP for $\gamma=\text{poly}(n)$ is also intractable for quantum computers. Up to date, the lack of efficient classical or quantum algorithms suggests that $\gamma$-SVP remains intractable for polynomial approximation factors. To evaluate algorithmic capabilities on solving $\gamma$-SVP, the approximation factors of SVP solvers are being considered. The approximation factor $\gamma$ of an SVP solver is defined as a ratio of the length of the returned vector $v$ from the algorithm and the length of the shortest non-zero vector of a lattice $\lambda_1$, $\gamma=\frac{\|v\|}{\lambda_1}$.

\section{Fixed-angle QAOA and Fixed-angle CM-QAOA Parameter pretraining}\label{sec:paramPretraining}
Because the angles are determined in advance, the classical optimization part of the QAOA algorithm is not performed. As a result, Fixed-angle (CM-)QAOA not only leverages a need to use hybrid classical-quantum setting, but it also significantly simplifies time analysis. In the standard QAOA, a major factor that complicated the time complexity analysis is a need to estimate number of optimizer iterations until convergence is reached. Experimentally observing the convergence for small instances cannot be extrapolated further as QAOA is vulnerable to barren plateaus \cite{McClean2018,Wang2021} which results in exponentially vanishing gradient and the rate at which this happens and its susceptibility to noise is difficult to determine. Fixed-angle (CM-)QAOA avoids this problem by omitting the optimization subroutine and, as a result, extrapolations of its performance to higher instance sizes are better justified. Due to lack of the optimization subroutine, the performance of Fixed-angle (CM-)QAOA is guaranteed to be worse than the performance of QAOA under the assumption that an optimizer can find the optimal QAOA angles. However, due to noise-induced barren plateaus and expensiveness of finding the optimal angles, this comparison is not expected to remain true as size of the instances grow to an extent that an optimizer is unable to find a reasonable minimum due to the aforementioned problems. We are inclined to conclude that in such cases the Fixed-angle (CM-)QAOA can overperform QAOA because Fixed-angle (CM-)QAOA would in this case provide a justification on the angles $\bm{\beta},\bm{\gamma}$ compared to QAOA which might with high probability return highly non-optimal angles. As an alternative, one might use the angles suggested by Fixed-angle (CM-)QAOA as a starting point for QAOA to pre-train the optimisation \cite{Shaydulin2019MultistartMF}. In this paper we experimentally determine the optimal angles for SVP instances that are of sizes implementable in our quantum emulations and at the same time which require a solution which cannot be guessed with a greater than $2^{-n}$ probability. This ensures that the determined angles are not biased towards a particular solution and the method of generating the experimental dataset is explained in details in \cref{sec:appGenRandom}.
\subsection{Parameter pretraining for the SVP problem}
We assume exponential time complexity of QAOA in solving the exact SVP problem. However, it is not theoretically ruled out that QAOA would solve SVP in sub-exponential time, but in this case the security of post-quantum cryptographic lattice-based proposals would be seriously threatened. Such a scenario as unlikely and we proceed to fit a function of the form $f(m)=e^{-am+b}$ to the results of Fixed-angle (CM-)QAOA. We generate dataset of $q$-ary lattices following a procedure described in \cref{sec:appGenRandom}. To pretrain the angles of Fixed-angle (CM-)QAOA we search for the optimal angles that minimise a variable $a$ when $f(m)$ is fit to pairs $$P^{m}(\bm{\beta},\bm{\gamma}):=\left(y_m=\begin{array}{c}
\text{averaged output of Fixed-angle (CM-)QAOA over }\\
m\text{ dimensional instances with angles }\bm{\beta},\bm{\gamma}
\end{array}, m=\begin{array}{c}
\text{dimension}\\
\text{of $q$-ary lattices}
\end{array}\right)$$
Suppose you are interested in estimation of the running time of Fixed-angle (CM-)QAOA with $m_{\text{target}}$ dimensional lattices from a lattice class $\bm{\mathcal{L}}$. Choose $m_{\text{start}}<m_{\text{end}}<<m_{\text{target}}$ so that you have technical capabilities to search for the optimal angles of Fixed-angle (CM-)QAOA up to $m_{\text{end}}$ dimensional instances, either on a quantum architecture or by running an emulation on a classical computer. The pretraining for the SVP problem for the lattice class $\bm{\mathcal{L}}$ is done in the following two steps:
\begin{enumerate}
	\item Generate datasets $\bm{\mathcal{L}}^{m_{\text{start}}}_{\text{train}},\bm{\mathcal{L}}^{m_{\text{start}+1}}_{\text{train}}\dots,\bm{\mathcal{L}}^{m_{\text{end}}}_{\text{train}}$ (Details in \cref{sec:appGenRandom})
	\item Use an optimiser to find the optimal angles $\bm{\beta}^\text{opt},\bm{\gamma}^\text{opt}$ that minimise a cost function $c_\text{train}(\bm{P}_{\text{train}}(\bm{\beta},\bm{\gamma}))$  that guides the optimiser to find the best possible Fixed-angle (CM-)QAOA results on the training set $\bm{P}_\text{train}^{m_{\text{i}}}(\bm{\beta},\bm{\gamma})=\left\{P_\text{train}^{m_{\text{start}}}(\bm{\beta},\bm{\gamma}),P_\text{train}^{m_{\text{start}+1}}(\bm{\beta},\bm{\gamma}),\dots,P_\text{train}^{m_{\text{end}}}(\bm{\beta},\bm{\gamma})\right\}$ where $P_\text{train}^{m_{\text{i}}}(\bm{\beta},\bm{\gamma})$ is a pair
	 $$P_\text{train}^{m_{\text{i}}}(\bm{\beta},\bm{\gamma})=\left(\begin{array}{c}
		\text{averaged output of Fixed-angle (CM-)QAOA over }\\
		\bm{\mathcal{L}}^{m_i}_{\text{train}}\text{ with angles }\bm{\beta},\bm{\gamma}
	\end{array}, m_i\right)$$
	The possible choices for $c_\text{train}(\cdot)$ that we found to perform particularly well are presented in \cref{sec:appCalcBestFit_costFn}.
	\item Return $\bm{\beta}^\text{opt},\bm{\gamma}^\text{opt}$ as the optimal pretrained angles and $a_\text{train}, b_\text{train}$ where $e^{-a_\text{train}m+b_\text{train}}$ is the best-fit (see \cref{sec:appCalcBestFit_expFit}) to the pairs $\bm{P}_\text{train}(\bm{\beta}^\text{opt},\bm{\gamma}^\text{opt})$.
\end{enumerate}

\subsection{Scalability Rationale for Fixed-angle (CM-)QAOA}
\milos{A fundamental question is whether fixed angles determined on small instances can remain effective as system size grows. While larger SVP instances have exponentially more degrees of freedom, Fixed-angle (CM-)QAOA with constant number of angles offers several key advantages: (I.) It bypasses optimization entirely, avoiding barren plateaus that make full QAOA intractable at scale. (II.) Our extrapolation methodology provides very pessimistic security bounds rather than performance claims. (III.) The structured nature of optimization problems may enable transfer of learned atterns across scales as was demonstrated in \cite{Boulebnane:2021sda,Boulebnane:2022ace,ParamConcentrations,Brando2018ForFC,9605328}. (IV.) Our experimental validation shows consistent trends across more than doubled problem sizes (training up to 10 dimensions, testing up to 22 dimensions), suggesting solid evidence.}

\section{Constraining the search space with CM-QAOA and Fixed-angle CM-QAOA Algorithm}\label{sec:constrainingWithFixQAOA}
A zero vector is not a valid solution to the SVP although it is included in the search space by using mappings as in \cite{notSoAdiabatic} or \cite{vqaSvpMilos}.  Although the latter work proposed also a different mapping that penalised the zero vector and made it a non-optimal solution, essentially reducing the probability that QAOA would return it, the mapping is not efficient as a linear number of qubits needs to be introduced in the dimension of a lattice. If the mappings that do not penalise zero vector solution are used, there is a unique state $\ket{\bm{\zeta}}$ that is the eigenstate of cost Hamiltonian $\hamiltonian_C$ corresponding to the zero lattice vector, i.e. $H_C\ket{\bm{\zeta}}=\bm{0}$.  We introduce a constrained mixing unitary $\mathcal{U}_M^{CM}$, that imposes a constraint that $\ket{\bm{\zeta}}=\ket{\zeta_1\zeta_2\dots\zeta_n}$ is not a feasible solution with which the alternating time evolutions under $\hamiltonian_C$ and $\mathcal{U}_M^{CM}$ leave the amplitude of $\ket{\bm{\zeta}}$ invariant. We define $\mathcal{U}_M^{CM}$ as a modification of a standard sum of Pauli$_X$ mixer, $\mathcal{U}_M^{CM}(\beta):=\sum_{i=1}^n \text{CTRL}\left(\overline{\zeta_i}\right)\text{-}R_{X_{1+(i\mod (n-1))}}(\beta)$ where $\text{CTRL}\left(\overline{\zeta_i}\right)\text{-}R_{X_{j}}(\beta)$ is a Pauli$_X$ rotation applied on $j$-th qubit controlled by $i$-th qubit. The control is triggered by the $i$-th qubit being in a state $\ket{0}$ if $\zeta_i=1$ and by state $\ket{1}$ if $\zeta_i=0$.

\begin{figure}[h!]
	\centering
	\vspace{-2em}
	\[
	\scalemath{0.7}{
		\Qcircuit @C=0.7em @R=0.6em {
			&\gate{X^{\zeta_1}} & \ctrl{1} & \gate{X^{\zeta_1}}& \qw & \qw & \qw & \qw & \gate{R_x(\beta)} & \qw& \qw\\
			&\qw & \gate{R_x(\beta)} & \gate{X^{\zeta_2}}& \ctrl{1} & \gate{X^{\zeta_2}} & \qw & \qw  & \qw& \qw & \qw\\
			&\qw & \qw & \qw & \gate{R_x(\beta)} & \gate{X^{\zeta_3}} &\ctrl{1} & \gate{X^{\zeta_3}} & \qw& \qw & \qw\\
			&\qw & \qw & \qw & \qw & \qw & \gate{R_x(\beta)} & \gate{X^{\zeta_4}} & \ctrl{-3} & \gate{X^{\zeta_4}} & \qw
		}
	}
	\]
	\caption{$\beta$-time evolution with 4-qubit mixing unitary $\mathcal{U}_M^{CM}$}
	\label{fig:groversSearch}
\end{figure}
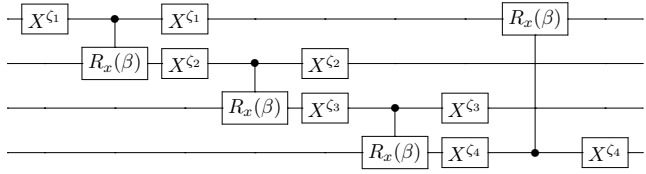
\begin{prop}\label{prop:probInvariant}
	The QAOA that uses $\mathcal{U}_M^{CM}$ as a mixing unitary leaves the amplitude of $\ket{\bm{\zeta}}$ invariant.
\end{prop}
\begin{proof}
	Let $P_{\bm{\zeta}}=\ket{\bm{\zeta}}\bra{\bm{\zeta}}$ be a projective measurement operator and denote $\ket{s}$ the starting state of QAOA (usually $\ket{s}=H^{\otimes n}\ket{\bm{0}}$) and $\ket{\psi_{\text{QAOA}}}$ the resulting state of $p$-layer QAOA. We proceed to show that $\braket{s|P_{\zeta}|s}=\braket{\psi_{\text{QAOA}}|P_{\zeta}|\psi_{\text{QAOA}}}$, i.e. that the amplitude of $\ket{\bm{\zeta}}$ is invariant to the run of QAOA. Observe that $\ket{\bm{\zeta}}$ is an eigenvector of $\mathcal{U}_M^{CM}$ with eigenvalue $1$ as by construction, no CTRL-$X$ gates are triggered. Consider an action of a single QAOA layer on a state $\ket{\bm{\zeta}}$:
	\begin{align*}
	e^{i\gamma_j\hamiltonian_C}\mathcal{U}_M^{CM}\ket{\bm{\zeta}}&=e^{i\gamma_j\hamiltonian_C}\ket{\bm{\zeta}}\\
	&=\sum_{k=0}^\infty\frac{1}{k!}\left[i\gamma_j\hamiltonian_C\right]^k \ket{\bm{\zeta}}\\
	&=\ket{\bm{\zeta}}\text{\ \ \ \ \ \ \ \ \ \ \ \ \ \ \ \ \ \ \ \ \ because $\hamiltonian_C\ket{\bm{\zeta}}=\bm{0}$}
	\end{align*}
	Then
	\begin{align*}
		\braket{\psi_{\text{QAOA}}|P_{\zeta}|\psi_{\text{QAOA}}}&=\bra{s}\prod_{j=1}^{p}e^{i\gamma_j\hamiltonian_C}\mathcal{U}_M^{CM}(\beta_j)\ket{\bm{\zeta}}\bra{\bm{\zeta}}\prod_{j=1}^{p}\mathcal{U}_M^{CM}(\beta_{p-j+1})e^{-i\gamma_{p-j+1}\hamiltonian_C}\ket{s}\\
		&=\braket{s|\bm{\zeta}}\braket{\bm{\zeta}|s}\\
		&=\braket{s|P_{\zeta}|s}
	\end{align*}
\end{proof}


\section{Solving the Exact-SVP}\label{sec:exactSVP}
Given an $n$-dimensional SVP instance, \cite{vqaSvpMilos} shows that using the mapping of Equation (\ref{eq:svpformulation2}), on average it is sufficient to allow for integer ranges $|x_i|\leq n$. Therefore, each coefficient $x_i$ on average requires $O(\log n)$ logical qubits. Therefore, following this approach one would on average need $O(n\log n)$ logical qubits to solve $n$-dimensional SVP instance. This is the same number of qubits as it is required by other NISQ approaches \cite{notSoAdiabatic,JCLM21,vqaSvpMilos}. However, by using the constrained mixing unitary $\mathcal{U}_M^{CM}$, Proposition \ref{prop:probInvariant} guarantees that the probability of obtaining the zero vector solution is $2^{-O(n\log n)}$ which is negligible. The previous approaches are expected to increase this probability. Although the magnification of this probability has not been rigorously studied (with the exception of small-scale experiments up to $n=6$ by \cite{JCLM21,notSoAdiabatic}), we conjecture that the magnification of the probability of returning zero vector would cause the approaches to be unscalable to larger lattice dimensions. Because in these approaches the zero vector represents the global optimum of the QAOA optimization routine, improving the QAOA performance would directly increase chance of calculated an invalid SVP solution. The experiments in this paper are performed by classical emulation with FastVQA \cite{FastVQA} and LattiQ \cite{LattiQ} libraries\footnote{We acknowledge GNU Parallel \cite{tange_2024_13957646} for distributing the computation amongst the computational nodes.}.

\subsection{Experimental Approach}\label{sec:expResults}
In this section we experimentally estimate and extrapolate upper bound on time complexity of QAOA-based SVP solver. We use the fact that QAOA performs at least as good as Fixed-angle QAOA and consequently an extrapolation on Fixed-angle QAOA-based SVP solver performance provides an upper bound on time complexity for solving SVP via QAOA. 
Our upper bound which we find below does not rely on any assumptions related to complexity of running an optimisation and its achieved optimiser performance or the presence of barren plateaus. Our estimated time complexity upper bound on QAOA-based SVP solver is therefore the time complexity of Fixed-angle QAOA based SVP solver.
The goals of the experiment which follows below are to:
\begin{enumerate}
	\item Test if the angles found by parameter pretraining (\Cref{sec:paramPretraining}) may be used in Fixed-angle (CM-)QAOA SVP solvers in larger dimensions than those that were used for training.
	\item Find the best exponential fit to time complexity of Fixed-angle QAOA and Fixed-angle CM-QAOA that serves as an upperbound for VQAs for SVP.
	\item Compare the performance of Fixed-angle (CM-)QAOA with the optimised (CM-)QAOA approach, with the random guess and with Grover's Algorithm performance.	
\end{enumerate}

We considered three different algorithm depths $p\in\{1,2,3\}$. For each depth (plotted as a separate graph of \Cref{fig:uniqueSVPResults}) the same dataset of SVP instances of dimensions 4 to 22 (with 100 different instances for each dimension) was used that was generated by the approach explained in \cref{sec:appGenRandom}. When mapping to a Hamiltonian formulation\footnote{\milos{For this we used the mapping by Equation \ref{eq:svpformulation2} in \Cref{sec:prelimSvpVqas}.}} a relaxed approach was followed, where 1 qubit was used to encode each of the coefficients $x_i$ (see more in \cref{sec:prelimSvpVqas}). In the actual attack on the SVP $O(\log n)$ qubits would be used to encode each of the coefficient. However, such a setting would considerably restrict the possible lattice sizes that can be classically emulated and experimented with. Therefore, in our experiments we do not solve the actual SVP but a relaxed version with a smaller search space which is not guaranteed to include the actual non-zero shortest lattice vector. The success of the experiment is instead calculated as a probability of retrieving the shortest non-zero vector of the search space defined by the assignment of 1 qubit to each of the coefficients $x_i$.  The same approach was followed e.g. in \cite{vqaSvpMilos} in which the authors argue that while such an assignment of qubits restrict a search space so that the actual shortest non-zero lattice vector is no longer guaranteed to be included, it retains the structure of the problem which is enough for evaluation of performance of NISQ algorithms. The parameter pretraining was performed separately for each depth on 4 to 10 dimensional instances and the performance of the determined angles is plotted as a solid line in the graphs. The angles were then used to evaluate the performance of Fixed-angle QAOA and Fixed-angle CM-QAOA algorithms in dimensions 11 to 22, essentially testing the scalability of determined angles up to problem instances of more than a double of size of the training instances. The observed performances are plotted with a dashed line. The best exponential fit of $2^{-am+b}$, with $m$ being the lattice dimension, is then performed to estimate the asymptotic scalings of the two approaches and its exponential scaling factor $a$ is printed. To evaluate the performance of Fixed-angle QAOA and Fixed-angle CM-QAOA in relation to other standard approaches, the performance of brute-force, Grover's algorithm and the standard optimised versions of QAOA and CM-QAOA are plotted. Note that due to time limitations of running classical emulations of optimised variational algorithms, the performance of QAOA and CM-QAOA, which is given for reference, was evaluated only up to lattice dimension 10.

\subsection{The Observed Performance}

\begin{figure*}
	\begin{subfigure}[b]{0.45\textwidth}  
		\hspace{20pt}
		\begin{tikzpicture}[scale=1]
			\begin{axis}[
	ymode=log,
	xmode=linear,
	log basis y={2},
	xmin = 3.8, xmax = 22.2,
	ymin = 7.74624248844371e-08,  ymax = 9,
	xtick distance = 2,
	ytick distance = 4,
	grid = both,
	minor tick num = 1,
	major grid style = {lightgray},
	minor grid style = {lightgray!25},
	width = \textwidth,
	height = \textwidth,
	xlabel = {Lattice basis dimension},
	ylabel = {$\log_{2}$(Overlap with solution)},
	x label style={at={(axis description cs:0.5,-0.05)},anchor=north},
	y label style={at={(axis description cs:-0.13,.5)},anchor=south},
	legend style={at={(1.4, 0.5)},anchor=west, legend cell align=left}, 
	]

	\addlegendimage{empty legend}
	\addlegendentry{\textbf{Fixed-angle CM-QAOA}}
	
	\addplot [line width=1pt, blue]
	plot [error bars/.cd, y dir = both, y explicit,
	error bar style={line width=0.1pt},
	error mark options={
		rotate=90,
		mark size=4pt,
		line width=0.1pt
	}]
	table[
	smooth,
	x expr={\thisrow{fixed_used_in_training}==0?nan:\thisrow{x}},
	y index=1,
	header=true,
	comment chars={21}
	] {chapters/figures/p_experiment_data/p1_experiments_scaling.dat};
	\addlegendentry{training}
		
	\addplot [line width=1pt, very thick,dash pattern={on 10pt off 2pt on 5pt off 2pt}, blue]
	plot [error bars/.cd, y dir = both, y explicit,
	error bar style={line width=0.1pt},
	error mark options={
		rotate=90,
		mark size=4pt,
		line width=0.1pt
	}]
	table[
	smooth,
	x expr={\thisrow{fixed_used_in_training2}==1?nan:\thisrow{x}},
	y index=1,
	header=true,
    y error plus expr={\thisrow{cm_overlap_std+}-\thisrow{cm_overlap_sv}},
	y error minus expr={\thisrow{cm_overlap_std+}-\thisrow{cm_overlap_sv}},
	] {chapters/figures/p_experiment_data/p1_experiments_scaling.dat};
	
	\addlegendentry{extrapolation}
	\addlegendimage{empty legend}
	\addlegendentry{}
	\addlegendimage{empty legend}
	\addlegendentry{\textbf{Fixed-angle QAOA}}
	
	\addplot[line width=1pt, brown]
		plot [error bars/.cd, y dir = both, y explicit,
	error bar style={line width=0.1pt},
	error mark options={
		rotate=90,
		mark size=4pt,
		line width=0.1pt
	}]
	table[
	smooth,
	x expr={\thisrow{fixed_used_in_training}==0?nan:\thisrow{x}},
	y index=4,
	header=true,
	comment chars={21}
	] {chapters/figures/p_experiment_data/p1_experiments_scaling.dat};
		
		\addlegendentry{training}
		
		\addplot[line width=1pt, very thick,dash pattern={on 10pt off 2pt on 5pt off 2pt}, brown]
	plot [error bars/.cd, y dir = both, y explicit,
	error bar style={line width=0.1pt},
	error mark options={
		rotate=90,
		mark size=4pt,
		line width=0.1pt
	}]
	table[
	smooth,
	x expr={\thisrow{fixed_used_in_training2}==1?nan:\thisrow{x}},
	y index=4,
	header=true,
    y error plus expr={\thisrow{qaoa_nonpen_overlap_std+}-\thisrow{qaoa_nonpen_overlap_sv}},
	y error minus expr={\thisrow{qaoa_nonpen_overlap_std+}-\thisrow{qaoa_nonpen_overlap_sv}},
	] {chapters/figures/p_experiment_data/p1_experiments_scaling.dat};
	
	\addlegendentry{extrapolation}
	\addlegendimage{empty legend}
	\addlegendentry{}
			
	\addplot[line width=0.1pt, red]
	table[
	smooth,
	x index=0,
	y index=1,
	header=false,
	] {chapters/figures/line.dat};
	
	\addlegendentry{Random guess $2^{-n}$}
	
	\addplot[line width=0.1pt, teal]
	table[
	smooth,
	x index=0,
	y index=2,
	header=false,
	] {chapters/figures/line.dat};
	
	\addlegendentry{Grover's algorithm}
	
	\addplot [line width=1pt, yellow]
	plot [error bars/.cd, y dir = both, y explicit,
	error bar style={line width=0.1pt},
	error mark options={
		rotate=90,
		mark size=4pt,
		line width=0.1pt
	}]
	table[
	smooth,
	x index=0,
	y index=1,
	header=true,
	comment chars={21}
	] {chapters/figures/p_experiment_data/p1_opt_experiments_scaling.dat};
	
	\addlegendentry{CM-QAOA}
	
		\addplot[line width=1pt, orange]
	plot [error bars/.cd, y dir = both, y explicit,
	error bar style={line width=0.1pt},
	error mark options={
		rotate=90,
		mark size=4pt,
		line width=0.1pt
	}]
	table[
	smooth,
	x index=0,
	y index=4,
	header=true,
	comment chars={21}
	] {chapters/figures/p_experiment_data/p1_opt_experiments_scaling.dat};
	
		\addlegendentry{QAOA (unconstrained)}
	
	\addplot[line width=0.1pt, dashed, blue]
	table[
	x index=0,
	y index=7,
	header=true,
	] {chapters/figures/p_experiment_data/p1_experiments_scaling.dat};
	
	\addplot[line width=0.1pt, dashed, brown]
	table[
	x index=0,
	y index=8,
	header=true,
	] {chapters/figures/p_experiment_data/p1_experiments_scaling.dat};

\pgfplotstableread[col sep=space]{chapters/figures/p_experiment_data/p1_experiments_scaling.dat}\datatableA
\node [brown] at (140,-15.7) {\rotatebox{                 -30.22      }{\small $a=$\pgfplotstablegetelem{1}{alpha_qaoa}\of{\datatableA}\pgfplotsretval}};
\node [blue] at (140,-18.5) {\rotatebox{                 -30.22      }{\small $a=$\pgfplotstablegetelem{1}{alpha_cm}\of{\datatableA}\pgfplotsretval}};

\end{axis}
		\end{tikzpicture}
		\caption[]%
		{{p=1}}    
	\end{subfigure}
	
		\centering
	\begin{subfigure}[b]{0.45\textwidth}
		\centering
		\begin{tikzpicture}[scale=1]
			\begin{axis}[
	ymode=log,
	xmode=linear,
	log basis y={2},
	xmin = 3.8, xmax = 22.2,
	ymin = 7.74624248844371e-08,  ymax = 9,
	xtick distance = 2,
	ytick distance = 4,
	grid = both,
	minor tick num = 1,
	major grid style = {lightgray},
	minor grid style = {lightgray!25},
	width = \textwidth,
	height = \textwidth,
	xlabel = {Lattice basis dimension},
	ylabel = {$\log_{2}$(Overlap with solution)},
	x label style={at={(axis description cs:0.5,-0.05)},anchor=north},
	y label style={at={(axis description cs:-0.13,.5)},anchor=south},
	legend style={at={(0.5,2)},anchor=north,legend cell align=left}]

	
	\addplot [line width=1pt, blue]
	plot [error bars/.cd, y dir = both, y explicit,
	error bar style={line width=0.1pt},
	error mark options={
		rotate=90,
		mark size=4pt,
		line width=0.1pt
	}]
	table[
	smooth,
	x expr={\thisrow{fixed_used_in_training}==0?nan:\thisrow{x}},
	y index=1,
	header=true,
	comment chars={21}
	] {chapters/figures/p_experiment_data/p2_experiments_scaling.dat};
	\addplot [line width=1pt, very thick,dash pattern={on 10pt off 2pt on 5pt off 2pt}, blue]
	plot [error bars/.cd, y dir = both, y explicit,
	error bar style={line width=0.1pt},
	error mark options={
		rotate=90,
		mark size=4pt,
		line width=0.1pt
	}]
	table[
	smooth,
	x expr={\thisrow{fixed_used_in_training2}==1?nan:\thisrow{x}},
	y index=1,
	header=true,
	y error plus expr={\thisrow{cm_overlap_std+}-\thisrow{cm_overlap_sv}},
	y error minus expr={\thisrow{cm_overlap_std+}-\thisrow{cm_overlap_sv}},
	] {chapters/figures/p_experiment_data/p2_experiments_scaling.dat};
	
	\addplot[line width=1pt, brown]
	plot [error bars/.cd, y dir = both, y explicit,
	error bar style={line width=0.1pt},
	error mark options={
		rotate=90,
		mark size=4pt,
		line width=0.1pt
	}]
	table[
	smooth,
	x expr={\thisrow{fixed_used_in_training}==0?nan:\thisrow{x}},
	y index=4,
	header=true,
	comment chars={21}
	] {chapters/figures/p_experiment_data/p2_experiments_scaling.dat};
	
	\addplot[line width=1pt, very thick,dash pattern={on 10pt off 2pt on 5pt off 2pt}, brown]
	plot [error bars/.cd, y dir = both, y explicit,
	error bar style={line width=0.1pt},
	error mark options={
		rotate=90,
		mark size=4pt,
		line width=0.1pt
	}]
	table[
	smooth,
	x expr={\thisrow{fixed_used_in_training2}==1?nan:\thisrow{x}},
	y index=4,
	header=true,
	y error plus expr={\thisrow{qaoa_nonpen_overlap_std+}-\thisrow{qaoa_nonpen_overlap_sv}},
	y error minus expr={\thisrow{qaoa_nonpen_overlap_std+}-\thisrow{qaoa_nonpen_overlap_sv}},
	] {chapters/figures/p_experiment_data/p2_experiments_scaling.dat};

	\addplot[line width=0.1pt, red]
	table[
	smooth,
	x index=0,
	y index=1,
	header=false,
	] {chapters/figures/line.dat};
	
	\addplot[line width=0.1pt, teal]
	table[
	smooth,
	x index=0,
	y index=2,
	header=false,
	] {chapters/figures/line.dat};
	
	\addplot [line width=1pt, yellow]
	plot [error bars/.cd, y dir = both, y explicit,
	error bar style={line width=0.1pt},
	error mark options={
		rotate=90,
		mark size=4pt,
		line width=0.1pt
	}]
	table[
	smooth,
	x index=0,
	y index=1,
	header=true,
	comment chars={21}
	] {chapters/figures/p_experiment_data/p2_opt_experiments_scaling.dat};
	
	\addplot[line width=1pt, orange]
	plot [error bars/.cd, y dir = both, y explicit,
	error bar style={line width=0.1pt},
	error mark options={
		rotate=90,
		mark size=4pt,
		line width=0.1pt
	}]
	table[
	smooth,
	x index=0,
	y index=4,
	header=true,
	comment chars={21}
	] {chapters/figures/p_experiment_data/p2_opt_experiments_scaling.dat};
	
	\addplot[line width=0.1pt, dashed, blue]
	table[
	x index=0,
	y index=7,
	header=true,
	] {chapters/figures/p_experiment_data/p2_experiments_scaling.dat};
	
	\addplot[line width=0.1pt, dashed, brown]
	table[
	x index=0,
	y index=8,
	header=true,
	] {chapters/figures/p_experiment_data/p2_experiments_scaling.dat};
	
	\pgfplotstableread[col sep=space]{chapters/figures/p_experiment_data/p2_experiments_scaling.dat}\datatableA
	\node [brown] at (140,-14.7) {\rotatebox{                 -30.22      }{\small $a=$\pgfplotstablegetelem{1}{alpha_qaoa}\of{\datatableA}\pgfplotsretval}};
	\node [blue] at (140,-18.5) {\rotatebox{                 -30.22      }{\small $a=$\pgfplotstablegetelem{1}{alpha_cm}\of{\datatableA}\pgfplotsretval}};

	\legend{}
\end{axis}
		\end{tikzpicture}
		\caption[Ex2]%
		{{p=2}}    
	\end{subfigure}
	\begin{subfigure}[b]{0.45\textwidth}  
		\centering 
		\begin{tikzpicture}[scale=1]
			\begin{axis}[
	ymode=log,
	xmode=linear,
	log basis y={2},
	xmin = 3.8, xmax = 22.2,
	ymin = 7.74624248844371e-08,  ymax = 9,
	xtick distance = 2,
	ytick distance = 4,
	grid = both,
	minor tick num = 1,
	major grid style = {lightgray},
	minor grid style = {lightgray!25},
	width = \textwidth,
	height = \textwidth,
	xlabel = {Lattice basis dimension},
	ylabel = {$\log_{2}$(Overlap with solution)},
	x label style={at={(axis description cs:0.5,-0.05)},anchor=north},
	y label style={at={(axis description cs:-0.13,.5)},anchor=south},
	legend style={at={(0.5,2)},anchor=north,legend cell align=left}]

	
	\addplot [line width=1pt, blue]
	plot [error bars/.cd, y dir = both, y explicit,
	error bar style={line width=0.1pt},
	error mark options={
		rotate=90,
		mark size=4pt,
		line width=0.1pt
	}]
	table[
	smooth,
	x expr={\thisrow{fixed_used_in_training}==0?nan:\thisrow{x}},
	y index=1,
	header=true,
	comment chars={21}
	] {chapters/figures/p_experiment_data/p3_experiments_scaling.dat};
	\addplot [line width=1pt, very thick,dash pattern={on 10pt off 2pt on 5pt off 2pt}, blue]
	plot [error bars/.cd, y dir = both, y explicit,
	error bar style={line width=0.1pt},
	error mark options={
		rotate=90,
		mark size=4pt,
		line width=0.1pt
	}]
	table[
	smooth,
	x expr={\thisrow{fixed_used_in_training2}==1?nan:\thisrow{x}},
	y index=1,
	header=true,
	y error plus expr={\thisrow{cm_overlap_std+}-\thisrow{cm_overlap_sv}},
	y error minus expr={\thisrow{cm_overlap_std+}-\thisrow{cm_overlap_sv}},
	] {chapters/figures/p_experiment_data/p3_experiments_scaling.dat};
	
	\addplot[line width=1pt, brown]
		plot [error bars/.cd, y dir = both, y explicit,
	error bar style={line width=0.1pt},
	error mark options={
		rotate=90,
		mark size=4pt,
		line width=0.1pt
	}]
	table[
	smooth,
	x expr={\thisrow{fixed_used_in_training}==0?nan:\thisrow{x}},
	y index=4,
	header=true,
	comment chars={21}
	] {chapters/figures/p_experiment_data/p3_experiments_scaling.dat};
	
		\addplot[line width=1pt, very thick,dash pattern={on 10pt off 2pt on 5pt off 2pt}, brown]
	plot [error bars/.cd, y dir = both, y explicit,
	error bar style={line width=0.1pt},
	error mark options={
		rotate=90,
		mark size=4pt,
		line width=0.1pt
	}]
	table[
	smooth,
	x expr={\thisrow{fixed_used_in_training2}==1?nan:\thisrow{x}},
	y index=4,
	header=true,
	y error plus expr={\thisrow{qaoa_nonpen_overlap_std+}-\thisrow{qaoa_nonpen_overlap_sv}},
	y error minus expr={\thisrow{qaoa_nonpen_overlap_std+}-\thisrow{qaoa_nonpen_overlap_sv}},
	] {chapters/figures/p_experiment_data/p3_experiments_scaling.dat};

	\addplot[line width=0.1pt, red]
	table[
	smooth,
	x index=0,
	y index=1,
	header=false,
	] {chapters/figures/line.dat};
	
	\addplot[line width=0.1pt, teal]
	table[
	smooth,
	x index=0,
	y index=2,
	header=false,
	] {chapters/figures/line.dat};
	
	\addplot [line width=1pt, yellow]
	plot [error bars/.cd, y dir = both, y explicit,
	error bar style={line width=0.1pt},
	error mark options={
		rotate=90,
		mark size=4pt,
		line width=0.1pt
	}]
	table[
	smooth,
	x index=0,
	y index=1,
	header=true,
	comment chars={21}
	] {chapters/figures/p_experiment_data/p3_opt_experiments_scaling.dat};
	
		\addplot[line width=1pt, orange]
	plot [error bars/.cd, y dir = both, y explicit,
	error bar style={line width=0.1pt},
	error mark options={
		rotate=90,
		mark size=4pt,
		line width=0.1pt
	}]
	table[
	smooth,
	x index=0,
	y index=4,
	header=true,
	comment chars={21}
	] {chapters/figures/p_experiment_data/p3_opt_experiments_scaling.dat};
	
	\addplot[line width=0.1pt, dashed, blue]
	table[
	x index=0,
	y index=7,
	header=true,
	] {chapters/figures/p_experiment_data/p3_experiments_scaling.dat};
	
	\addplot[line width=0.1pt, dashed, brown]
	table[
	x index=0,
	y index=8,
	header=true,
	] {chapters/figures/p_experiment_data/p3_experiments_scaling.dat};

\pgfplotstableread[col sep=space]{chapters/figures/p_experiment_data/p3_experiments_scaling.dat}\datatableA
\node [brown] at (140,-13.7) {\rotatebox{                 -30.22      }{\small $a=$\pgfplotstablegetelem{1}{alpha_qaoa}\of{\datatableA}\pgfplotsretval}};
\node [blue] at (140,-18.5) {\rotatebox{                 -30.22      }{\small $a=$\pgfplotstablegetelem{1}{alpha_cm}\of{\datatableA}\pgfplotsretval}};

	\legend{}
\end{axis}
		\end{tikzpicture}
		\caption[]%
		{{p=3}}    
	\end{subfigure}
	\caption[]{{Experimental results for solving the Exact-SVP}} \label{fig:uniqueSVPResults}
\end{figure*}
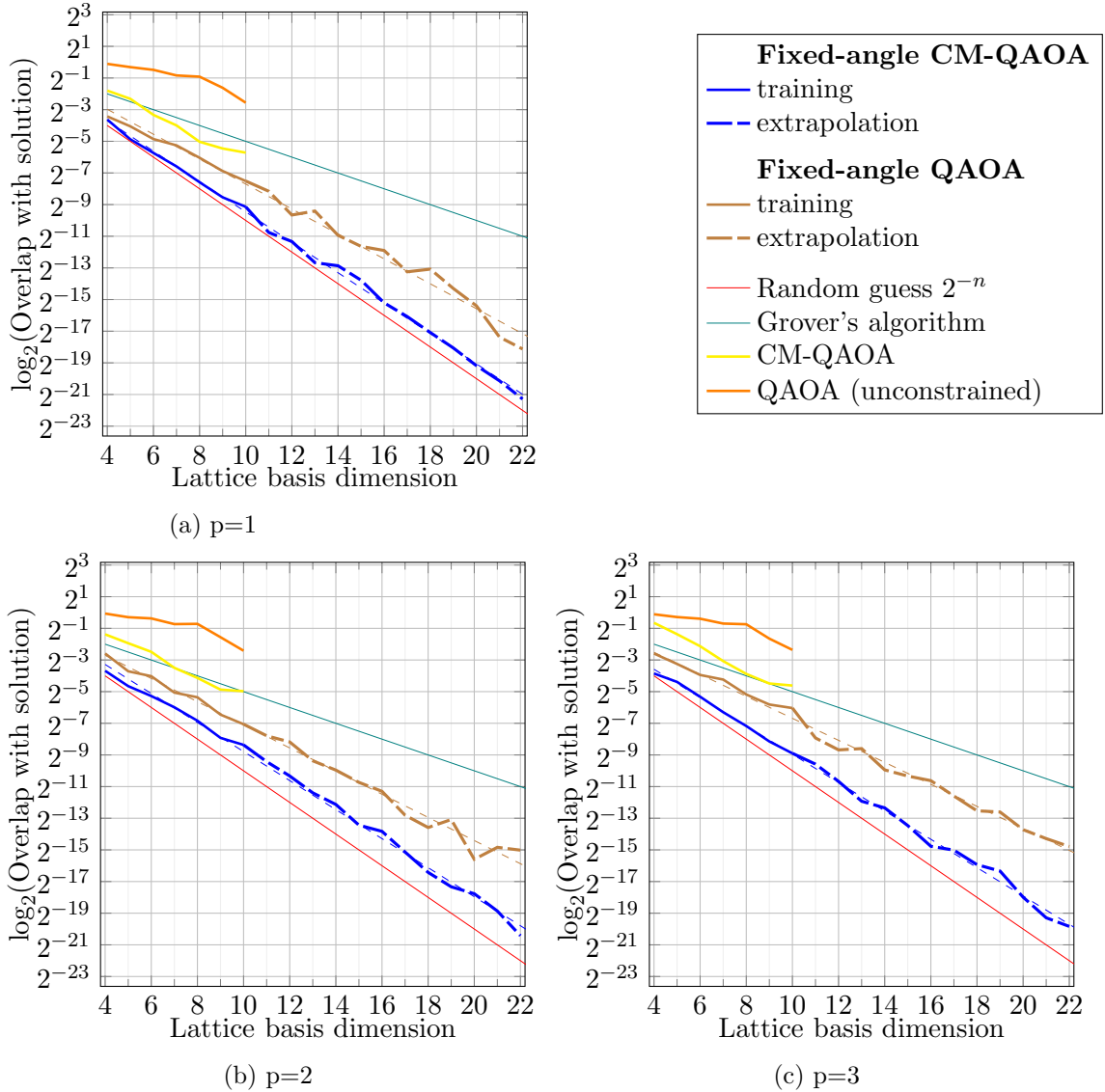

The results, plotted in \Cref{fig:uniqueSVPResults}, show that the asymptotic speedup of Fixed-angle QAOA and Fixed-angle CM-QAOA improves with the depth $p$ of the algorithms. The exponential factors $a$ are printed alongside the best fits of the form $2^{-am+b}$. For the largest depth $p=3$ considered in our experiments we have observed that Fixed-angle QAOA solves the problem in $2^{0.695m+O(m)}$ time and Fixed-angle CM-QAOA in $2^{0.895m+O(m)}$ time. We note that black-box Grover's algorithm runs in $2^{0.5m+O(m)}$ time (green line)\footnote{Because we assigned 1 qubit per dimension, one should substitute $m=\tfrac{3}{2}n\log n+O(\log n)$, with $n$ being lattice dimension, for comparison with classical SVP solvers. This follows from Theorem 1 of \cite{vqaSvpMilos}.}. It is important to note that while Fixed-angle QAOA with depth $p=3$ has a slightly worse accuracy than Grover's search, it is an algorithm with a polynomial depth that is relatively low when compared to Grover's algorithm whose depth has an exponential growth. To highlight the importance of finding a constant-depth NISQ algorithm we emphasise that to the best of our knowledge all the FT quantum SVP solver proposals either require implementation of black-box oracles (e.g. \cite{montanaro2018quantum,kuperberg2005svp,arora2017closelyrelated}) or assume qRAM \cite{lovenstein2017quantum} in case of quantum sieving algorithms. Their implementation on quantum hardware can be costly as it was shown e.g. for the Grover's SVP solver \cite{prokop2024groversoracleshortestvector} and use of error-correcting techniques further increase quantum resource demands. It is also evident that Fixed-angle QAOA has consistently performed better than Fixed-angle CM-QAOA despite that the latter has a strategy to avoid the zero-vector solution which the former does not have. We are of the opinion that with increasing $p\gg 3$ this would no longer be true and Fixed-angle CM-QAOA would outperform Fixed-angle QAOA\footnote{This expectation is based on the observation that Fixed-angle CM-QAOA introduces additional structure into the ansatz by explicitly avoiding the zero-vector solution, thereby focusing the optimisation on meaningful candidates. While this restriction can limit the expressivity of shallow circuits and thus hinder performance at low depths, we hypothesise that at larger depths, the increased expressivity of the ansatz will compensate for this constraint. In fact, the restricted search space becomes advantageous, as it excludes invalid solutions and may help steer the optimisation towards the shortest non-zero vector.}. A more detailed comparison between the two algorithms is in \Cref{sec:approxFactors} once their performance on Approximate SVP is studied in \Cref{sec:approxSVP}. The performances of the optimisation algorithms QAOA and CM-QAOA has been included for the first 10 dimensions for the reference.

\section{Approximate SVP solver}\label{sec:approxSVP}
While results in \cref{sec:expResults} evaluated performance of Fixed-angle QAOA and Fixed-angle CM-QAOA on the exact SVP problem, many practical post-quantum cryptography proposals rely on a relaxed version of SVP, the approximate SVP $\gamma$-SVP which asks for the shortest vector with the length of factor $\gamma$ times the exact length of the shortest non-zero lattice vector $\lambda_1$. These proposals (e.g. LWE \cite{regev2009lattices}, Ring-LWE \cite{lyubashevsky2010ideal} or Ajtai-Dwork \cite{ajtaiCryptosystemPolyApprox}) rely on conjectured hardness of solving $\gamma$-SVP with polynomial factor $\gamma=n^c$ for a constant $c$. The constant $c$ depends on the actual cryptographic proposal and its implementation. In this section we choose a few values of $c$ and demonstrate the performance of running Fixed-angle QAOA and Fixed-angle CM-QAOA on $n^c$-SVP. The results are useful for estimation of performance of variational quantum algorithms on $\gamma$-SVP problem which is of a greater importance compared to its exact version when quantum capabilities on real-world cryptographic instances are evaluated.

\subsection{Solving approximate SVP problem}
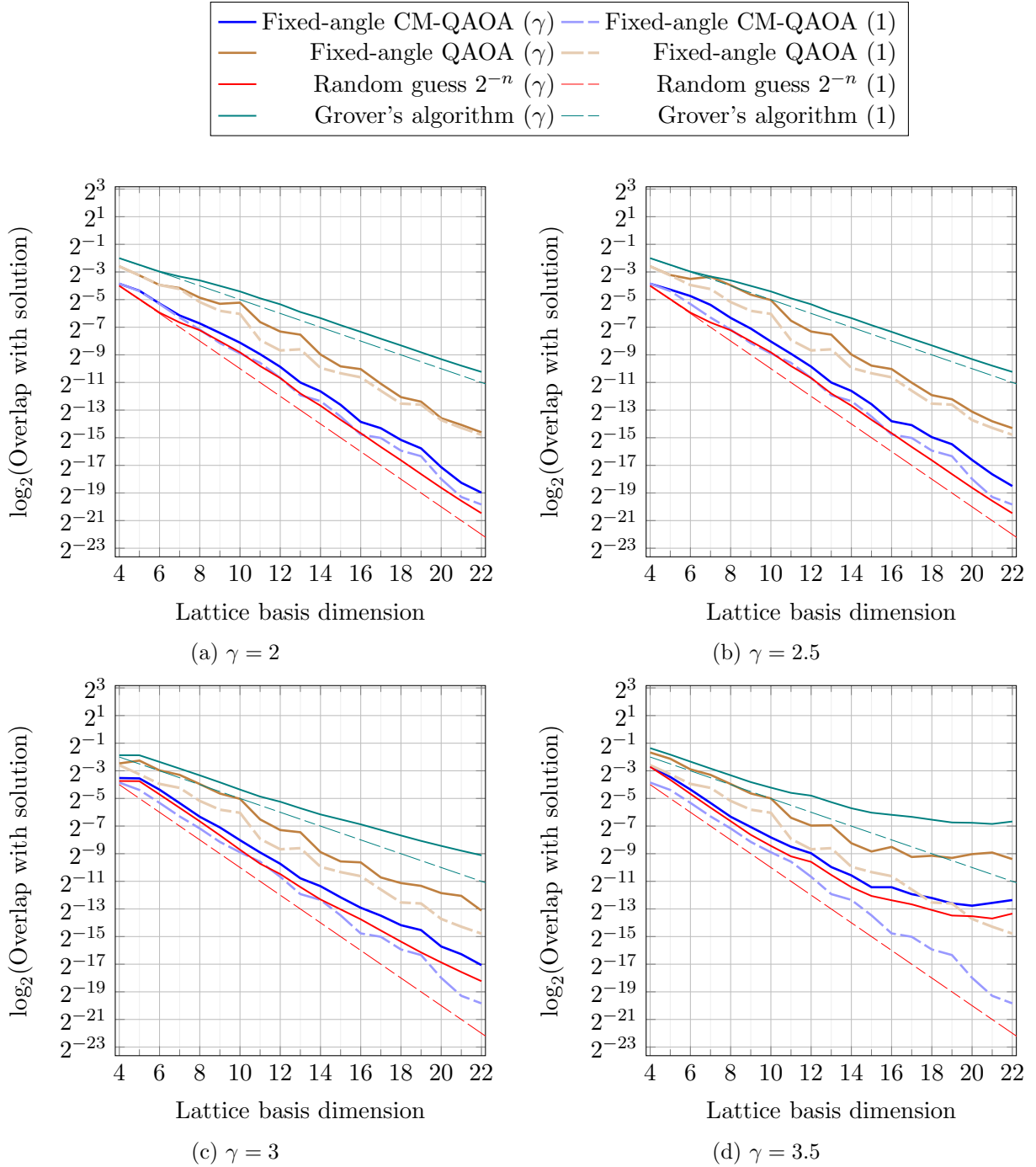
\begin{figure*}
	\centering
	\begin{subfigure}[b]{0.45\textwidth}
		\centering
		\begin{tikzpicture}[scale=1]
	\begin{axis}[
	ymode=log,
	xmode=linear,
	log basis y={2},
	xmin = 3.8, xmax = 22.2,
	ymin = 7.74624248844371e-08,  ymax = 9,
	xtick distance = 2,
	ytick distance = 4,
	grid = both,
	minor tick num = 1,
	major grid style = {lightgray},
	minor grid style = {lightgray!25},
	width = \textwidth,
	height = \textwidth,
	xlabel = {Lattice basis dimension},
	ylabel = {\(\log_{2}\)(Overlap with solution)},
	label style={font=\normalsize},
	tick label style={font=\normalsize},
	x label style={at={(axis description cs:0.5,-0.1)},anchor=north},
	y label style={at={(axis description cs:-0.2,.5)},anchor=south},
	legend style={at={(1.2,1.5)},anchor=north,legend cell align=right, legend columns=2}]
	
	\addplot [line width=1pt, blue]
	plot [error bars/.cd, y dir = both, y explicit,
	error bar style={line width=0.1pt},
	error mark options={
		rotate=90,
		mark size=4pt,
		line width=0.1pt
	}]
	table[
	smooth,
	y index=1,
	header=true,
	y error plus expr={\thisrow{cm_err}-\thisrow{cm_overlap_sv}},
	y error minus expr={\thisrow{cm_err}-\thisrow{cm_overlap_sv}},
	] {chapters/figures/p_experiment_data/approx_p3_experiments_scaling_2.dat};
	
	\addplot [line width=1pt, blue!40, dash pattern={on 10pt off 2pt on 5pt off 2pt}]
	plot [error bars/.cd, y dir = both, y explicit,
	error bar style={line width=0.1pt},
	error mark options={
		rotate=90,
		mark size=4pt,
		line width=0.1pt
	}]
	table[
	smooth,
	y index=1,
	header=true,
	] {chapters/figures/p_experiment_data/p3_experiments_scaling.dat};

	\addplot[line width=1pt, brown]
plot [error bars/.cd, y dir = both, y explicit,
error bar style={line width=0.1pt},
error mark options={
	rotate=90,
	mark size=4pt,
	line width=0.1pt
}]
table[
smooth,
y index=2,
header=true,
y error plus expr={\thisrow{qaoa_err}-\thisrow{qaoa_nonpen_overlap_sv}},
y error minus expr={\thisrow{qaoa_err}-\thisrow{qaoa_nonpen_overlap_sv}},
] {chapters/figures/p_experiment_data/approx_p3_experiments_scaling_2.dat};

\addplot[line width=1pt, very thick,dash pattern={on 10pt off 2pt on 5pt off 2pt}, brown!40]
plot [error bars/.cd, y dir = both, y explicit,
error bar style={line width=0.1pt},
error mark options={
	rotate=90,
	mark size=4pt,
	line width=0.1pt
}]
table[
smooth,
y index=4,
header=true,
] {chapters/figures/p_experiment_data/p3_experiments_scaling.dat};

	\addplot[line width=0.75pt, red]
	table[
	smooth,
	x index=0,
	y index=1,
	header=true,
	] {chapters/figures/p_experiment_data/approx_p3_experiments_scaling_2_line.dat};
	
	\addplot[line width=0.1pt, red, dash pattern={on 10pt off 2pt on 5pt off 2pt}]
	table[
	smooth,
	x index=0,
	y index=1,
	header=false,
	] {chapters/figures/line.dat};

	\addplot[line width=0.75pt, teal]
	table[
	smooth,
	x index=0,
	y index=2,
	header=true,
	] {chapters/figures/p_experiment_data/approx_p3_experiments_scaling_2_line.dat};
	
	\addplot[line width=0.1pt, teal, dash pattern={on 10pt off 2pt on 5pt off 2pt}]
	table[
	smooth,
	x index=0,
	y index=2,
	header=false,
	] {chapters/figures/line.dat};
	
	\legend{Fixed-angle CM-QAOA (\(\gamma\)), Fixed-angle CM-QAOA (1), Fixed-angle QAOA (\(\gamma\)), Fixed-angle QAOA (1), Random guess \(2^{-n}\) (\(\gamma\)), Random guess \(2^{-n}\) (1), Grover's algorithm (\(\gamma\)), Grover's algorithm (1)}
\end{axis}
		\end{tikzpicture}
		\caption[Ex2]%
		{{\(\gamma=2\)}}    
	\end{subfigure}
	\hspace{20pt}
	\begin{subfigure}[b]{0.45\textwidth}  
		\centering 
		\begin{tikzpicture}[scale=1]
			\begin{axis}[
	ymode=log,
	xmode=linear,
	log basis y={2},
	xmin = 3.8, xmax = 22.2,
	ymin = 7.74624248844371e-08,  ymax = 9,
	xtick distance = 2,
	ytick distance = 4,
	grid = both,
	minor tick num = 1,
	major grid style = {lightgray},
	minor grid style = {lightgray!25},
	width = \textwidth,
	height = \textwidth,
	xlabel = {Lattice basis dimension},
	ylabel = {\(\log_{2}\)(Overlap with solution)},
	label style={font=\normalsize},
	tick label style={font=\normalsize},
	x label style={at={(axis description cs:0.5,-0.1)},anchor=north},
	y label style={at={(axis description cs:-0.2,.5)},anchor=south},
	legend style={at={(0.5,2)},anchor=north,legend cell align=left}]
	
	\addplot [line width=1pt, blue]
	plot [error bars/.cd, y dir = both, y explicit,
	error bar style={line width=0.1pt},
	error mark options={
		rotate=90,
		mark size=4pt,
		line width=0.1pt
	}]
	table[
	smooth,
	y index=1,
	header=true,
	y error plus expr={\thisrow{cm_err}-\thisrow{cm_overlap_sv}},
	y error minus expr={\thisrow{cm_err}-\thisrow{cm_overlap_sv}},
	] {chapters/figures/p_experiment_data/approx_p3_experiments_scaling_2.5.dat};
	
	\addplot [line width=1pt, blue!40, dash pattern={on 10pt off 2pt on 5pt off 2pt}]
	plot [error bars/.cd, y dir = both, y explicit,
	error bar style={line width=0.1pt},
	error mark options={
		rotate=90,
		mark size=4pt,
		line width=0.1pt
	}]
	table[
	smooth,
	y index=1,
	header=true,
	] {chapters/figures/p_experiment_data/p3_experiments_scaling.dat};

	\addplot[line width=1pt, brown]
plot [error bars/.cd, y dir = both, y explicit,
error bar style={line width=0.1pt},
error mark options={
	rotate=90,
	mark size=4pt,
	line width=0.1pt
}]
table[
smooth,
y index=2,
header=true,
y error plus expr={\thisrow{qaoa_err}-\thisrow{qaoa_nonpen_overlap_sv}},
y error minus expr={\thisrow{qaoa_err}-\thisrow{qaoa_nonpen_overlap_sv}},
] {chapters/figures/p_experiment_data/approx_p3_experiments_scaling_2.5.dat};

\addplot[line width=1pt, very thick,dash pattern={on 10pt off 2pt on 5pt off 2pt}, brown!40]
plot [error bars/.cd, y dir = both, y explicit,
error bar style={line width=0.1pt},
error mark options={
	rotate=90,
	mark size=4pt,
	line width=0.1pt
}]
table[
smooth,
y index=4,
header=true,
] {chapters/figures/p_experiment_data/p3_experiments_scaling.dat};

	\addplot[line width=0.1pt, red, dash pattern={on 10pt off 2pt on 5pt off 2pt}]
table[
smooth,
x index=0,
y index=1,
header=false,
] {chapters/figures/line.dat};

	\addplot[line width=0.75pt, red]
	table[
	smooth,
	x index=0,
	y index=1,
	header=true,
	] {chapters/figures/p_experiment_data/approx_p3_experiments_scaling_2_line.dat};
	
		\addplot[line width=0.1pt, teal, dash pattern={on 10pt off 2pt on 5pt off 2pt}]
	table[
	smooth,
	x index=0,
	y index=2,
	header=false,
	] {chapters/figures/line.dat};
	
	\addplot[line width=0.75pt, teal]
	table[
	smooth,
	x index=0,
	y index=2,
	header=true,
	] {chapters/figures/p_experiment_data/approx_p3_experiments_scaling_2_line.dat};
		
	
	\legend{}
\end{axis}
		\end{tikzpicture}
		\caption[]%
		{{\(\gamma=2.5\)}}    
	\end{subfigure}
	
	\centering
	\begin{subfigure}[b]{0.45\textwidth}
		\centering
		\begin{tikzpicture}[scale=1]
			\begin{axis}[
	ymode=log,
	xmode=linear,
	log basis y={2},
	xmin = 3.8, xmax = 22.2,
	ymin = 7.74624248844371e-08,  ymax = 9,
	xtick distance = 2,
	ytick distance = 4,
	grid = both,
	minor tick num = 1,
	major grid style = {lightgray},
	minor grid style = {lightgray!25},
	width = \textwidth,
	height = \textwidth,
	xlabel = {Lattice basis dimension},
	ylabel = {\(\log_{2}\)(Overlap with solution)},
	label style={font=\normalsize},
	tick label style={font=\normalsize},
	x label style={at={(axis description cs:0.5,-0.1)},anchor=north},
	y label style={at={(axis description cs:-0.2,.5)},anchor=south},
	legend style={at={(0.5,2)},anchor=north,legend cell align=left}]
	
	\addplot [line width=1pt, blue]
	plot [error bars/.cd, y dir = both, y explicit,
	error bar style={line width=0.1pt},
	error mark options={
		rotate=90,
		mark size=4pt,
		line width=0.1pt
	}]
	table[
	smooth,
	y index=1,
	header=true,
	y error plus expr={\thisrow{cm_err}-\thisrow{cm_overlap_sv}},
	y error minus expr={\thisrow{cm_err}-\thisrow{cm_overlap_sv}},
	] {chapters/figures/p_experiment_data/approx_p3_experiments_scaling_3.dat};
	
	\addplot [line width=1pt, blue!40, dash pattern={on 10pt off 2pt on 5pt off 2pt}]
	plot [error bars/.cd, y dir = both, y explicit,
	error bar style={line width=0.1pt},
	error mark options={
		rotate=90,
		mark size=4pt,
		line width=0.1pt
	}]
	table[
	smooth,
	y index=1,
	header=true,
	] {chapters/figures/p_experiment_data/p3_experiments_scaling.dat};

	\addplot[line width=1pt, brown]
plot [error bars/.cd, y dir = both, y explicit,
error bar style={line width=0.1pt},
error mark options={
	rotate=90,
	mark size=4pt,
	line width=0.1pt
}]
table[
smooth,
y index=2,
header=true,
y error plus expr={\thisrow{qaoa_err}-\thisrow{qaoa_nonpen_overlap_sv}},
y error minus expr={\thisrow{qaoa_err}-\thisrow{qaoa_nonpen_overlap_sv}},
] {chapters/figures/p_experiment_data/approx_p3_experiments_scaling_3.dat};

\addplot[line width=1pt, very thick,dash pattern={on 10pt off 2pt on 5pt off 2pt}, brown!40]
plot [error bars/.cd, y dir = both, y explicit,
error bar style={line width=0.1pt},
error mark options={
	rotate=90,
	mark size=4pt,
	line width=0.1pt
}]
table[
smooth,
y index=4,
header=true,
] {chapters/figures/p_experiment_data/p3_experiments_scaling.dat};

	\addplot[line width=0.1pt, red, dash pattern={on 10pt off 2pt on 5pt off 2pt}]
table[
smooth,
x index=0,
y index=1,
header=false,
] {chapters/figures/line.dat};

	\addplot[line width=0.75pt, red]
	table[
	smooth,
	x index=0,
	y index=1,
	header=true,
	] {chapters/figures/p_experiment_data/approx_p3_experiments_scaling_3_line.dat};
	
		\addplot[line width=0.1pt, teal, dash pattern={on 10pt off 2pt on 5pt off 2pt}]
	table[
	smooth,
	x index=0,
	y index=2,
	header=false,
	] {chapters/figures/line.dat};
	
	\addplot[line width=0.75pt, teal]
	table[
	smooth,
	x index=0,
	y index=2,
	header=true,
	] {chapters/figures/p_experiment_data/approx_p3_experiments_scaling_3_line.dat};
		
	
	\legend{}
\end{axis}
		\end{tikzpicture}
		\caption[Ex2]%
		{{\(\gamma=3\)}}    
	\end{subfigure}
	\hspace{20pt}
	\begin{subfigure}[b]{0.45\textwidth}  
		\centering 
		\begin{tikzpicture}[scale=1]
			\begin{axis}[
	ymode=log,
	xmode=linear,
	log basis y={2},
	xmin = 3.8, xmax = 22.2,
	ymin = 7.74624248844371e-08,  ymax = 9,
	xtick distance = 2,
	ytick distance = 4,
	grid = both,
	minor tick num = 1,
	major grid style = {lightgray},
	minor grid style = {lightgray!25},
	width = \textwidth,
	height = \textwidth,
	xlabel = {Lattice basis dimension},
	ylabel = {\(\log_{2}\)(Overlap with solution)},
	label style={font=\normalsize},
	tick label style={font=\normalsize},
	x label style={at={(axis description cs:0.5,-0.1)},anchor=north},
	y label style={at={(axis description cs:-0.2,.5)},anchor=south},
	legend style={at={(0.5,2)},anchor=north,legend cell align=left}]
	
	\addplot [line width=1pt, blue]
	plot [error bars/.cd, y dir = both, y explicit,
	error bar style={line width=0.1pt},
	error mark options={
		rotate=90,
		mark size=4pt,
		line width=0.1pt
	}]
	table[
	smooth,
	y index=1,
	header=true,
	y error plus expr={\thisrow{cm_err}-\thisrow{cm_overlap_sv}},
	y error minus expr={\thisrow{cm_err}-\thisrow{cm_overlap_sv}},
	] {chapters/figures/p_experiment_data/approx_p3_experiments_scaling_3.5.dat};
	
	\addplot [line width=1pt, blue!40, dash pattern={on 10pt off 2pt on 5pt off 2pt}]
	plot [error bars/.cd, y dir = both, y explicit,
	error bar style={line width=0.1pt},
	error mark options={
		rotate=90,
		mark size=4pt,
		line width=0.1pt
	}]
	table[
	smooth,
	y index=1,
	header=true,
	] {chapters/figures/p_experiment_data/p3_experiments_scaling.dat};

	\addplot[line width=1pt, brown]
	plot [error bars/.cd, y dir = both, y explicit,
	error bar style={line width=0.1pt},
	error mark options={
		rotate=90,
		mark size=4pt,
		line width=0.1pt
	}]
	table[
	smooth,
	y index=2,
	header=true,
	y error plus expr={\thisrow{qaoa_err}-\thisrow{qaoa_nonpen_overlap_sv}},
	y error minus expr={\thisrow{qaoa_err}-\thisrow{qaoa_nonpen_overlap_sv}},
	] {chapters/figures/p_experiment_data/approx_p3_experiments_scaling_3.5.dat};
	
	\addplot[line width=1pt, very thick,dash pattern={on 10pt off 2pt on 5pt off 2pt}, brown!40]
	plot [error bars/.cd, y dir = both, y explicit,
	error bar style={line width=0.1pt},
	error mark options={
		rotate=90,
		mark size=4pt,
		line width=0.1pt
	}]
	table[
	smooth,
	y index=4,
	header=true,
	] {chapters/figures/p_experiment_data/p3_experiments_scaling.dat};
	
	\addplot[line width=0.1pt, red, dash pattern={on 10pt off 2pt on 5pt off 2pt}]
	table[
	smooth,
	x index=0,
	y index=1,
	header=false,
	] {chapters/figures/line.dat};
	
	\addplot[line width=0.75pt, red]
	table[
	smooth,
	x index=0,
	y index=1,
	header=true,
	] {chapters/figures/p_experiment_data/approx_p3_experiments_scaling_3.5_line.dat};
	
	\addplot[line width=0.1pt, teal, dash pattern={on 10pt off 2pt on 5pt off 2pt}]
	table[
	smooth,
	x index=0,
	y index=2,
	header=false,
	] {chapters/figures/line.dat};
	
	\addplot[line width=0.75pt, teal]
	table[
	smooth,
	x index=0,
	y index=2,
	header=true,
	] {chapters/figures/p_experiment_data/approx_p3_experiments_scaling_3.5_line.dat};
	
	
	\legend{}
\end{axis}
		\end{tikzpicture}
		\caption[]%
		{{\(\gamma=3.5\)}}    
	\end{subfigure}
	
	\caption[]{{Experimental results for solving the $\gamma$-SVP}. (\(\gamma\) is the approximate version corresponding to \(\gamma\) in the captions and (1) is the exact, \(\gamma=1\) version provided for reference. Depth 3 Fixed-angle (CM-)QAOA algorithms are used.} \label{fig:exactAndApproxP3}
\end{figure*}

To evaluate the scaling of Fixed-angle QAOA and Fixed-angle CM-QAOA on $\gamma$-SVP we have experimented with $\gamma\in\{2,3,4\}$\milos{, see Figure \ref{fig:exactAndApproxP3}}. The same experimental dataset as in \cref{sec:expResults} was used, the search space hence remains the same. However, the solution space is enlarged in the $\gamma$-SVP problem. Let $\tau$ be number of solutions (i.e. non-zero lattice vectors of the length $\leq\gamma\lambda_1$) for our experimental set of dimension $n$, then for the $n$-dimensional $\gamma$-SVP the random guess approach (plotted in red) and Grover's algorithm (plotted in green) have $\tau2^{n}$ and $\lceil\frac{\pi}{4}\rceil\sqrt{\frac{2^n}{\tau}}$ time complexities respectively. For each dimension both Fixed-angle QAOA and Fixed-angle CM-QAOA have been run and the results are plotted in brown and blue respectively.
The dashed lines correspond to the exact, \(\gamma=1\) scenario, whereas the solid lines correspond to the \(\gamma>1\) scenario as indicated in subplots' captions.
The angles have been reused from the training performed in \cref{sec:expResults}.

Empirical observation of the results of \Cref{fig:exactAndApproxP3} reveals that, under our experimental conditions, both Fixed-angle QAOA and Fixed-angle CM-QAOA do not provide any considerable advantage for solving approximate \(\gamma\)-SVP compared to their performance on the exact SVP of \Cref{fig:uniqueSVPResults}. One of the likely reasons is that pre-training the QAOA on the exact SVP leads to a trained algorithm that learns to find the global minimum, rather than having a more spreaded distribution over the lower eigen-spectrum of the problem Hamiltonian. The future work is needed to test this hypothesis, which would in fact, further validate our method for QAOA pretraining for SVP. It is also possible that this observation is influenced by our choice of a small constant \(\gamma\), which keeps the approximate solution space relatively close to the exact one. However, larger values of \(\gamma\) may not correspond to practically relevant cryptographic regimes, as most proposals rely on \(\gamma = \mathrm{poly}(n)\).

\subsection{Approximation factors of Fixed-angle QAOA and Fixed-angle CM-QAOA}\label{sec:approxFactors}

To better compare Fixed-angle QAOA and Fixed-angle CM-QAOA we focus on so-called approximation factors. Approximation factor is a metric that allows for comparison of SVP solvers based on their capabilites on solving the approximate SVP (see \Cref{sec:prelimApproxSVP}). Variational quantum algorithms produce quantum states which are superpositions of different outcomes with a hope that a probability of sampling an outcome that is a solution to a problem is high enough. \cref{sec:evalApproxFactor} explains a technique to calculate the approximation factor from an output of QAOA and \cref{sec:expResultsApproxFactors} presents the experimental results.

\subsubsection{Evaluating approximation factor of a variational quantum SVP solver}\label{sec:evalApproxFactor}
In order to evaluate the approximation factor of the algorithms we calculate the expected length of the vector sampled from such an output superposition. A result of a variational quantum algorithm is a quantum state $\ket{\psi}$ such that $\ket{\psi}=\sum_{i\in\{0,1\}^{\otimes\text{rows}(\hamiltonian)}} \sqrt{p_i} \ket{i}$ where $p_i$ are probabilities of sampling the computational states $\ket{i}$.  Since we are clasically emulating variational quantum algorithms, we have direct access to the states $\ket{i}$ and their corresponding probabilities $p_i$. We can therefore efficiently post-process the result to get probabilities of sampling vectors corresponding to the successive minima $\lambda_0,\lambda_1,\dots$ as this post-processing only involves clustering the vectors into equivalence classes where an equivalence relation is defined by equality of the lengths of the vectors. The corresponding probabilities related to these classes are then summed probabilities of sampling the vectors inside the clusters. Let $p_{\lambda_i}$ be a probability of sampling a vector of length $\lambda_i$.  Given that we do not sample a zero vector (otherwise we say that the algorithm fails and this happens with probability $p_{\lambda_0}$), the expected length of the output vector $\mathit{l}\in\mathcal{L}$ by running an algorithm $\mathcal{A}$ on the lattice $\lattice$ is
$$
\mathbb{E}_{\mathcal{A}}\left[\min_{\mathit{l}\in\lattice}||l||_2\right]=\sum_{i>0}\lambda_ip_{\lambda_i}
$$
Suppose we run $\mathcal{A}$ and sample the output $k$ times so that we retrieved $k$ (not necessarily distinct) lattice vectors and chose the smallest non-zero one as the result of the algorithm. To evaluate expected length of the algorithm's output vector we introduce variables
\[X_j=\lambda_i\iff \text{a vector of length $\lambda_i$ was drawn on $j$-th draw}\]
The expected length of the output vector $\mathit{l}$ by running $\mathcal{A}$ $k$-times (denoted by $\mathcal{A}^k$) on the lattice $\lattice$ is then
\begin{align}
	\mathbb{E}_{\mathcal{A}^k}\left[\min_{\mathit{l}\in\lattice}||l||_2\right]&=\sum_{i>0}\lambda_i\probability\left(\left(\forall j\text{ s.t. } 0\leq j < k: X_j \geq \lambda_i\text{ or }X_j = 0\right)\wedge\left(\exists j\text{ s.t. } 0\leq j < k: X_j=\lambda_i \right)\right)\nonumber\\
	&=\sum_{i>0}\lambda_i\biggl(\probability\left( \forall j\text{ s.t. } 0\leq j < k: X_j\geq \lambda_i\text{ or } X_j = 0\right)\nonumber\\
	&\text{\ \ \ \ \ \ \ \ \ \ \ \ \ \ }-\probability\left( \forall j\text{ s.t. } 0\leq j < k: X_j > \lambda_i\right)-\probability\left( \forall j\text{ s.t. } 0\leq j < k: X_j =0\right)\biggr)\nonumber\\
	&=\sum_{i>0}\lambda_i\left[\left(\left(\sum_{j>i}p_{\lambda_j}\right)+p_{\lambda_0}+p_{\lambda_i}\right)^k-\left(\sum_{j>i}p_{\lambda_j}\right)^k-p_{\lambda_0}^k\right]\label{eq:expected_length_qaoa_k}
\end{align}
Note that $\mathcal{A}^k$ has a probability $p_{\lambda_0}^k$ of failure that happens when we sample zero vector after each of $k$-runs of evaluation of $\mathcal{A}$. The Equation \ref{eq:expected_length_qaoa_k} gives us the expected output of $\mathcal{A}^k$ if such failure has not happened.

\subsubsection{Experimental results}\label{sec:expResultsApproxFactors}

\begin{figure}[H]
	\centering
	\begin{tikzpicture}[scale=0.4]
		\begin{axis}[
	ymode=linear,
	xmode=linear,
	xmin = 3.8, xmax = 22.2,
	ymin = 0,  ymax = 750,
	xtick distance = 2,
	ytick distance = 100,
	grid = both,
	minor tick num = 1,
	major grid style = {lightgray},
	minor grid style = {lightgray!25},
	width = \textwidth,
	height = \textwidth,
	xlabel = {Lattice basis dimension},
	ylabel = {Approximation Factor},
	label style={font=\huge},
	tick label style={font=\huge},
	x label style={at={(axis description cs:0.5,-0.05)},anchor=north},
	y label style={at={(axis description cs:-0.1,.5)},anchor=south},
	legend style={at={(0,0.5)},anchor=west,legend cell align=left, font=\huge},
	legend pos=north west]
	
	\addplot [line width=1pt, blue]
	plot [error bars/.cd, y dir = both, y explicit,
	error bar style={line width=0.1pt},
	error mark options={
		rotate=90,
		mark size=4pt,
		line width=0.1pt
	}]
	table[
	smooth,
	y index=1,
	header=true,
	] {chapters/figures/approx_factors.dat};
	
	\addplot [line width=1pt, red]
	plot [error bars/.cd, y dir = both, y explicit,
	error bar style={line width=0.1pt},
	error mark options={
		rotate=90,
		mark size=4pt,
		line width=0.1pt
	}]
	table[
	smooth,
	y index=2,
	header=true,
	] {chapters/figures/approx_factors.dat};
	
	\addplot [line width=1pt, very thick,dash pattern={on 10pt off 2pt on 5pt off 2pt}, blue]
	plot [error bars/.cd, y dir = both, y explicit,
	error bar style={line width=0.1pt},
	error mark options={
		rotate=90,
		mark size=4pt,
		line width=0.1pt
	}]
	table[
	smooth,
	y index=3,
	header=true,
	] {chapters/figures/approx_factors.dat};
	
	\addplot [line width=1pt, very thick,dash pattern={on 10pt off 2pt on 5pt off 2pt}, red]
	plot [error bars/.cd, y dir = both, y explicit,
	error bar style={line width=0.1pt},
	error mark options={
		rotate=90,
		mark size=4pt,
		line width=0.1pt
	}]
	table[
	smooth,
	y index=4,
	header=true,
	] {chapters/figures/approx_factors.dat};
	
	\legend{Fixed-angle CM\_QAOA, Fixed-angle QAOA, Best fit $0.06n^{2.92}$, Best fit $0.06n^{3.04}$ }
\end{axis}
	\end{tikzpicture}
	\caption{Approximate factors of Fixed-angle QAOA and Fixed-angle CM-QAOA algorithms}
	\label{fig:approxFactors}
\end{figure}
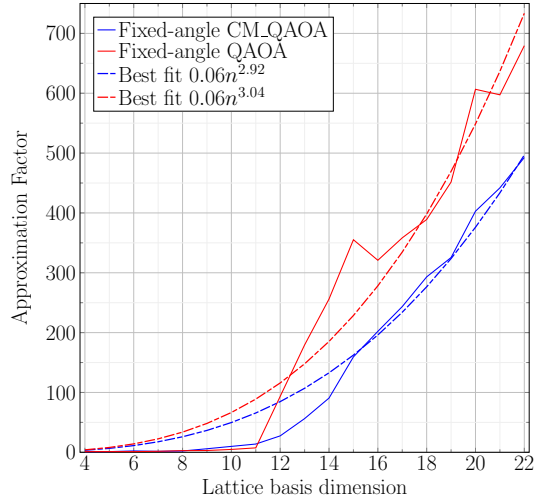

We used \cref{eq:expected_length_qaoa_k} to express expected lenghts of the sampled vectors from our algorithms, and by dividing with the actual lengths of the shortest non-zero lattice vectors we calculated the approximation factors of Fixed-angle QAOA and Fixed-angle CM-QAOA algorithms. The \cref{eq:expected_length_qaoa_k} converges to the exact expected output as the number of samples $k$ increases. We have hardcoded $k=5000$ in our experiments which is a great overestimation on the value of $k$ for which the output already converged accurately to a precision of a few decimal places, therefore it was a reasonable choice and the presented results are a very accurate estimation of the expected outputs. For each lattice dimension we have used the same SVP instance datasets as in the experiments on exact SVP in \cref{sec:expResults} and the results are plotted in \cref{fig:approxFactors}.

\begin{figure*}
\centering
\input{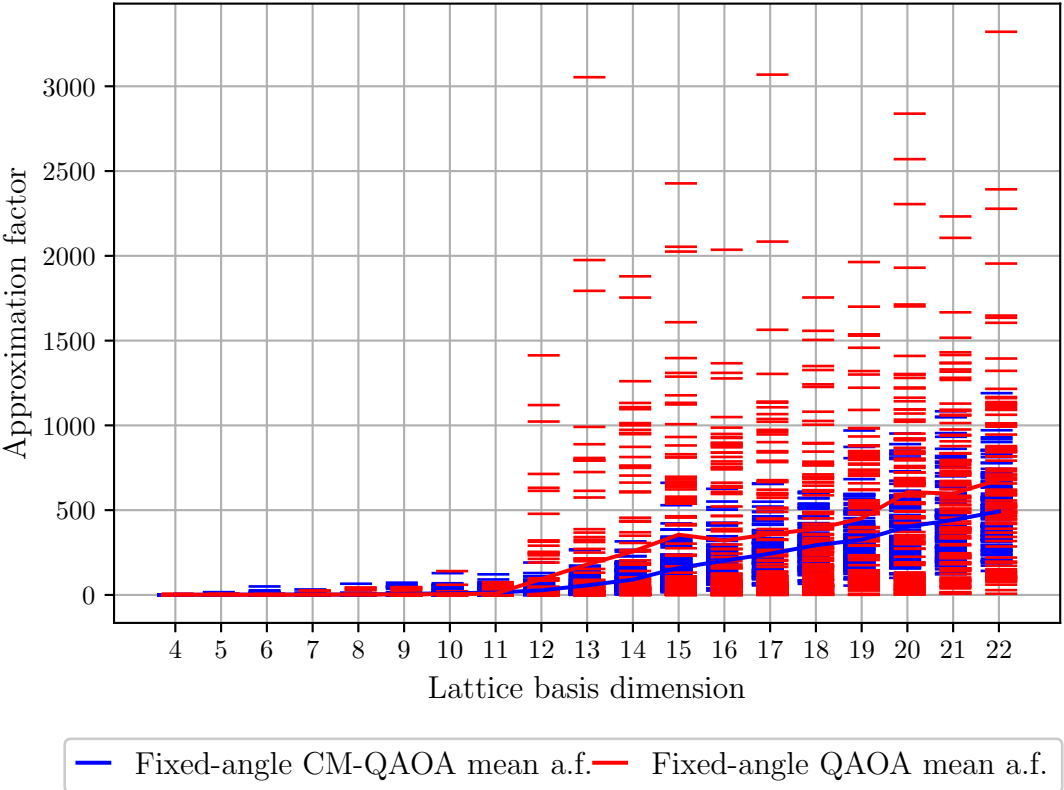}
\caption{The distributions of approximate factors obtained from solving 100 instances using Fixed-angle CM-QAOA and Fixed-angle QAOA. The solid lines represent the average values, which are identical to those in \Cref{fig:approxFactors}. This figure provides additional insight by illustrating the difference in variability of the distributions of the approximate factors produced by the two algorithms. The individual approximation factors are plotted with short horizontal lines, with small horizontal shifts to improve the readability.}\label{fig:approxFactorsOneByOne}
\end{figure*}

The approximations factors for Fixed-angle CM-QAOA were observed on average to be lower than the approximation factors for Fixed-angle QAOA. This is in contrast, but not in contradiction, to findings from \Cref{sec:expResults} in which Fixed-angle QAOA performed better even though it was not penalised and included the undesired zero vector in the solution space. The reasons are apparent after a closer inspection of \Cref{fig:approxFactorsOneByOne}
which extends the plot \Cref{fig:approxFactors} by plotting the approximation factors of 100 individual experimental instances before an average is calculated. We can see that the approximation factors of Fixed-angle CM-QAOA are distributed much more densely around their mean than the approximation factors of Fixed-angle QAOA, which have shown relatively a very large spread. This means that the performance of Fixed-angle CM-QAOA is more stable and can be more reliably extrapolated than the performance of (non-penalised) Fixed-angle QAOA. This suggests that the use of Fixed-angle CM-QAOA is a better alternative for further experiments on estimating the performance of QAOA SVP solvers as it provides more stable and more densely distributed output within certain approximation factors with little outliers (compared to Fixed-angle QAOA whose approximation factors distribution is significantly skewed and contains many outliers). These advantages of Fixed-angle CM-QAOA are at the cost of a slightly worse performance than Fixed-angle QAOA on average.

%

\section{Conclusion}\label{sec:cmQaoaConclusions}
\textbf{Summary}. This paper proposed pre-training of QAOA angles on solving the SVP (Fixed-angle QAOA). We have demonstrated that for angles trained on lattices up to dimension 10, the angles still showed a stable performance on lattice instances of more than double dimension, the 22 dimensional instances. We expect the trend to continue, i.e. that the performance would persist even with larger lattice instances as no limiting optimiser-related phenomena are involved. We have two reasons for this conjecture. Firstly, the method of pre-training QAOA angles has been validated in \cite{Boulebnane:2022ace} for solving another type of structured combinatorial optimisation problem, k-SAT, and has been shown to match performance of analytically determined upper-bound on the performance of QAOA k-SAT solver. Secondly, we have been pessimistic about the choice of best-fit function used to evaluate the performance of Fixed-angle (CM-)QAOA method. Although we have chosen to fit the exponential function of the form $2^{-\alpha n+b}$, the actual algorithm performance may also be sub-exponential (but not super exponential). In that case we provide even looser upper bounds on the actual performance of (CM-)QAOA to solve SVP, and the actual scalings may be even more optimistic than our conservatively determined ones\footnote{\milos{With the assumption of exponentially small overlap of Fixed-angle (CM)QAOA with the solution, our method does not propose an efficient way to attack post-quantum cryptosystems. Rather, it investigates practical scales of the speed-up offered by NIST architectures.}}. We have also proposed a novel technique to avoid converging with QAOA towards the trivial, zero vector solution. Previous approaches have either ignored this problem \cite{notSoAdiabatic,JCLM21}, which was demonstrated to be a sufficient approach for very small lattice dimensions or proposed to introduce a penalty \cite{vqaSvpMilos} which introduces additional qubits\footnote{Although it introduces only $n-2$ extra qubits for an $n$ dimensional lattice, this can mean a non-negligible difference for low-qubit NISQ era architectures.}. The new techniques makes advantage of the fact that the zero vector solution is known a priori and uses a generalisation of QAOA, the Quantum Alternative Operator Ansatz, with a specifically designed mixing unitary to avoid convergence towards it. We have called the resulting algorithm CM-QAOA (controlled-mixer QAOA) and its pre-trained version Fixed-angle CM-QAOA, and experimentally compared their performances with a standard QAOA and (Fixed-angle QAOA algorithm). We found that despite constraining the search space with (Fixed-angle) CM-QAOA, QAOA was still more efficient to find the solutions to SVP (i.e. in this case the first excited states of problem Hamiltonians). We conjecture this is caused by a usage of a mixing unitary used in (Fixed-angle) CM-QAOA algorithm that mixes slower and requires greater algorithmic depth $p$ to match the performance of (Fixed-angle) QAOA. We expect that once this extra depth $p$ is reached, the performance would be qualitatively better than the non constrained one. Moreover, this underperformance of (fixed-angle) CM-QAOA does not undermine its usefulness: (i) in contrast to (Fixed-angle) QAOA, it is an algorithm that provably returns the SVP solution in the infinite limit of depth $p$ of the algorithm, (ii) it is better suited for extrapolations of the performance to higher dimension due to having more stable performance as shown in \Cref{sec:approxFactors}.

The experiments have been run for both exact SVP, where the viable solution is only a shortest non-zero lattice vector, and approximate \(\gamma\)-SVP which also considers approximations to the shortest non-zero lattice vector up to a polynomial factor \(\gamma\). The extrapolations suggest \(2^{0.695m+O(m)}\) and \(2^{0.895m+O(m)}\) time scalings for Fixed-angle QAOA and Fixed-angle CM-QAOA respectively on their ability to solve the exact SVP.

The work has also compared Fixed-angle QAOA and Fixed-angle CM-QAOA by their approximation factors. It has been found that even though Fixed-angle QAOA has performed better on the exact SVP on average, it has higher approximation factors on average, meaning that it performs worse on approximate \(\gamma\)-SVP than Fixed-angle CM-QAOA. The reason was revealed after closer inspection by findings in \Cref{fig:approxFactorsOneByOne} where individual instance runs were being analysed. It has been found that Fixed-angle QAOA has much bigger and more skewed spread over Hamiltonian eigen-spectrum in its output and that the reason behind its high approximation factors are the outliers (where Fixed-angle QAOA did not perform well at all) that shift the average. However, the 25\% quantile of the data for Fixed-angle QAOA in \Cref{fig:approxFactorsOneByOne} is considerably smaller than that of Fixed-angle CM-QAOA, indicating that Fixed-angle QAOA more consistently finds shorter solutions. On the other hand, Fixed-angle CM-QAOA has much more dense output distribution over still relatively very short solutions with no outliers. It therefore seems like a more viable candidate for the extrapolation of the performance of QAOA based SVP solvers if stability and reliability is of the major concern, even at the expense of a slightly worse performance (i.e. less tight upper bound on the performance of QAOA based SVP).


\petros{Finally, we note that in our approach we quantify the time-complexity by the inverse of the success probability. One could try to do a similar analysis for Grover's algorithm, where instead of using sufficient queries to solve the problem deterministically, they could choose to ``stop earlier'' leading to success with some (potentially small) probability, and then repeat the whole process sufficient times. However, we note that since the advantage comes from the amplitude amplification of sequential oracle applications, any attempts to terminate the amplification earlier (and then repeat the whole process) would lead to a higher overall query count -- related with the difficulty in parallelizing Grover's algorithm. It follows that the correct comparison with Grover-based approaches is the one we made here.}
\\

\noindent \textbf{Limitations}. The asymptotic scalings have been observed up to 22 dimensional lattice instances (while being trained on lattices up to dimension 10). It is not clear if and to what extent would the performance remain stable on $>22$ dimensional lattices, although, as we have explained above, we expect that the observed performance extrapolates well. 
We further, note that the experiments in this paper were classically emulated and assumed an ideal, noiseless scenario. The effects of noise would affect our observed performance of Fixed-angle (CM-)QAOA algorithms, although to a less extent than during the noisy execution of the optimised \mbox{(CM-)}QAOA algorithm. Nevertheless, the effects of noise need to be considered if one considers practical implications of our results. We finally stress that Fixed-angle (CM-)QAOA algorithms are expected to perform worse than the optimised (CM-)QAOA algorithms because the optimisation part is missing and the algorithms are made \textit{less powerful}\footnote{Although this may not always be true. An example are scenarious when the optimisation part of (CM-)QAOA fails, e.g. due to the effects of barren plateaux.}. Therefore performance of Fixed-angle (CM-)QAOA algorithms in general upper bounds the performance of (CM-)QAOA algorithms on their ability to solve SVP. However, we do not know how tight this bound is, i.e. it is not clear how much better would (CM-)QAOA algorithms perform given the extrapolated performances of Fixed-angle (CM-)QAOA algorithms.
\\

\noindent \textbf{Future directions}. Further research is required to validate our observations and conjectured expectations about the ability to use the pre-trained QAOA angles for solving much larger SVP instances. A possible direction might be proving an analytic formula for the bound of using Fixed-angle QAOA as it has been done in \cite{Boulebnane:2022ace} for the k-SAT problem. It is also not clear how the pre-training of QAOA work in the practical, noisy scenarios. The performance may be worsened both during the training (up to a failure to determine the optimal angles) or during the use of the determined angles in solving larger dimensional SVP instances. We note however, that compared to noisy, real-world execution of QAOA, the noisy execution of Fixed-angle (CM-)QAOA is expected to be much more robust to the effects of noise as the noise-induced barren plateaux \cite{Wang2021} are not present.
Furthermore, the experiments on \(\gamma\)-SVP have not demonstrated any considerable improvement better than solving exact SVP. We conjecture that this is because the Fixed-angle (CM-)QAOA algorithms learn to find the ground state of the problem Hamiltonian while ignoring other low energy solutions. Therefore the output distribution demonstrates no significant spread over the lower energy eigen-spectrum of the problem Hamiltonian. If our conjecture is true, the evidence for suitability of QAOA angle pretraining for SVP is even stronger. Further research is required to verify the conjecture by comparing the presented angle pre-training methods with pre-training where low energy solutions (approximations of SVP) are also being favoured.
We also note that experiments for different algorithmic depths (greater than $p=3$ used in this paper), or considerations of the implications of the findings for Learning with Errors \cite{regev2009lattices} problem are other possible continuations of our work.


\section*{Acknowledgments}
PW acknowledges support by EPSRC grants EP/T001062/1, EP/X026167/1, EP/T026715/1, STFC grant ST/W006537/1 and Edinburgh-Rice Strategic Collaboration Awards and MP acknowledges support by EPSRC DTP studentship grant EP/T517811/1.


\appendix
\section*{\large Appendices} 
\Cref{sec:appGenRandom} introduces an efficient way to generate SVP instances for QAOA and CM-QAOA angle pretraining. \Cref{sec:AppcmQAOApreTrainingEssentials} gives details about the choices of cost functions used in QAOA and CM-QAOA angle pretraining.
\section{Generating random lattice instances}\label{sec:appGenRandom}
We choose \(\varsigma\in\mathbb{Z}_q^m\) to be a random vector which will become a solution of an instance generated. We proceed to generate lattice basis \(\basis\) such that \(\basis\varsigma=\lambda_1\) is the shortest non-zero lattice vector. \(q\) is expected to be a prime number. The larger the \(q\), the more instances can be generated with the given solution \(\varsigma\). However, too large \(q\) can lead to numerical instabilities in the optimisation process\footnote{These numerical instabilities can be always fixed by increasing numerical precisions, e.g. using the GNU MP Bignum Library \cite{gmp}, they unnecessarily add to computational complexity of experiments.} to find the solution to SVP. Since this approach is not used with a strict requirement of adhering to some lattice hardness results, but rather used as a method for VQAs pre-training, we suggest to keep \(q\) only as large as it provides sufficient number of instances. What is of a particular importance about this method is that it allows for a choice of a spectral gap\footnote{Which is the difference between energy of ground state and first excited state \(\lvert k_1-0\rvert=k_1\).} and the \textit{first-excited state gap}\footnote{Which is the difference between energy of first and second excited states \(\lvert k_1-k_2 \rvert\).}. These gaps are important for the performance of VQAs as the spectral gap is directly correlated with the time evolution in Adiabatic Theorem which is in turn the theoretical foundation of QAOA, and the \textit{first-excited state gap} is interesting from lattice perspectives as it dictates the difference of the lengths of the shortest and the second shortest non-zero vectors in a lattice. Also very importantly, this approach allows to set the gaps (which are chosen by setting the magnitudes \(k_1,\dots,k_m\)) that correspond to the cryptographic relevant lattice constructions of much higher ranks. Consequently, with careful setting of the parameters, it is possible to use this approach to construct small lattice instances with (some of the) essential security parameters of the large lattice instances.

The sampling procedure of \(\basis\) follows these steps:

\begin{enumerate}
	\item \label{item:first}Choose a length of the future solution \(\lambda_1\in\mZ\), i.e. of the future shortest non-zero vector of \(B\). There are two options:
	\begin{enumerate}
		\item Set the value of \(\lambda_1\) manually if it is intended to generate lattice with the exact pre-set non-zero shortest vector length
		\item Sample length \(\lambda_1>0\) of the future shortest non-zero vector of \(B\) according to a user-defined distribution (e.g. uniform distribution over integers on the interval \([a,b]\) for \(a>0\) and \(b\) reasonably large)
	\end{enumerate}
	\item Denote \(e_1:=\varsigma\) and generate orthonormal basis \(e_2,\dots,e_{m}\) for \(e_1^\perp\subset\mathbb{Z}^{(m-1)\times m}\).\footnote{It is convenient to use Gram-Schmidt Orthogonalisation to find the orthonormal basis perpendicular to \(e_1\).}
	\item Let \(k_1:=\frac{\lambda_1}{\|\varsigma\|}\) and choose \(k_2\gg\frac{\lambda_1}{\|\varsigma\|^2}\), the length of the second shortest non-zero vector in the lattice and randomly sample \(k_3,\dots,k_m\geq k_2\), the eigenvalues associated with the eigenspace perpendicular to the first two shortest non-zero lattice vectors
	\item \label{item:last}Let \[P:=\begin{pmatrix} \vdots & \vdots & & \vdots \\
		\varsigma & e_2 & \cdots & e_{m} \\
		\vdots & \vdots  & & \vdots \end{pmatrix}\]
		This allows us to calculate \(\basis\) from its eigendecomposition \[\basis:=P\begin{pmatrix}
		\frac{\lambda_1}{|\varsigma|} & 0 & \cdots & 0\\
		0 & k_2 &  \cdots & 0\\
		\vdots  & \vdots &\ddots &  0 \\
		0 & 0 & \cdots  & k_m
	\end{pmatrix}P^{-1}\]
\end{enumerate}
\begin{prop}
	\(\varsigma\) is a unique solution corresponding to the shortest non-zero vector of lattice basis \(\basis\).
\end{prop}
\begin{proof}[{Proof that steps \Cref{item:first}--\Cref{item:last} generate the desired basis \(\basis\)}] By eigendecomposition of \(\basis\) we know that \(\|B\varsigma\|=\left\|\frac{\lambda_1}{|\varsigma|}\varsigma\right\|=\lambda_1\). Let \(x\in\mathbb{Z}_q^m\setminus 0^{\otimes m}, x \neq \varsigma\) be a non-zero vector that does not correspond to the solution of SVP on the lattice with basis \(\basis\). Then
	\begin{align*}
		\|\basis x\|^2&=x^T\basis^T\basis x\\
		&=\left(\sum_{1\leq i\leq m}\langle x, e_i \rangle e_i^T\basis^T\right)\left(\sum_{1\leq i\leq m}\langle x, e_i \rangle \basis e_i\right)\\
		&=\left(\sum_{1\leq i\leq m}\langle x, e_i \rangle k_i e_i^T\right)\left(\sum_{1\leq i\leq m}\langle x, e_i \rangle k_ie_i\right)\\
		&\ \ \text{because of orthogonality of \(\{e_1,\dots,e_m\}\)}\\
		&=\sum_{1\leq i\leq m}\langle x, e_i \rangle^2\langle e_i, e_i \rangle k_i^2\\
		&> \sum_{1\leq i\leq m}\langle x, e_i \rangle^2\langle e_i, e_i \rangle \lambda_1^2\\
		&=\langle x, \varsigma \rangle^2\|\varsigma\|^2\frac{\lambda_1^2}{\|\varsigma\|^2}+\lambda_1^2\sum_{2\leq i\leq m}\langle x, e_i \rangle^2 \\
		&=\langle x, \varsigma \rangle^2\lambda_1^2+\lambda_1^2 (\|x\|^2-\langle x,\varsigma\rangle^2)\\
		&=\lambda_1^2 \|x\|^2
	\end{align*}
	This implies that \(\|\basis x\| > \lambda_1\|x\| > \lambda_1 = \|\basis\varsigma\|\) where we used the fact that \(\|x\| \geq 1\) as \(x\in\mathbb{Z}_q^m\setminus 0^{\otimes m}\). This concludes the proof.
\end{proof}

\section{Fixed-angle QAOA and Fixed-angle CM-QAOA Pre-Training Essentials}\label{sec:AppcmQAOApreTrainingEssentials}
Pre-training of Fixed-angle QAOA and Fixed-angle CM-QAOA requires a few non obvious choices for the cost function being used and the procedure to estimate the algorithmic performance given some particular trial set of angles. In \Cref{sec:appCalcBestFit_costFn} we introduce the cost function that we found to perform well during our pre-training method. It involves estimation of an exponent of a best-fit, which is then discussed in \Cref{sec:appCalcBestFit_expFit}.
\subsection{The cost function used in parameter pretraining}\label{sec:appCalcBestFit_costFn}
The parameter pretraining of Fixed-angle (CM-)QAOA  finds the optimal angles \(\bm{\beta}^\text{opt},\bm{\gamma}^\text{opt}\) that result in a \textit{good} performance over the training dataset \(\bm{\mathcal{L}}^{m_{\text{start}}}_{\text{train}},\bm{\mathcal{L}}^{m_{\text{start}+1}}_{\text{train}}\dots,\bm{\mathcal{L}}^{m_{\text{end}}}_{\text{train}}\). The \textit{good} performance can be understood in terms of different metrics which are encoded in the cost function \(c_\text{train}(\bm{P}_{\text{train}}(\bm{\beta},\bm{\gamma}))\) used in the optimisation. We have found two implementations of the cost function,  \(c^1_\text{train}(\bm{P}_{\text{train}}(\bm{\beta},\bm{\gamma}))\) and \(c^2_\text{train}(\bm{P}_{\text{train}}(\bm{\beta},\bm{\gamma}))\) (each of which has two variations of being deterministic or randomised), that worked particularly well with the goal of finding Fixed-angle (CM-)QAOA angles that provide reliable extrapolation to higher lattice dimensions which also provides a significant advantage in solving SVP in the extrapolated dimensions.

\subsubsection{\(c^1_\text{train}\): Maximising the Exponent of the Speedup}
The first approach is to maximise the exponent of the speedup\footnote{We assume the exponential decay of success probability of solving SVP as the dimension \(m\) increases. This is consistent with the expectation that VQAs would not outperform best known quantum fault-tolerant SVP solvers, which solve SVP in exponential time.} which is equivalent to minimising the exponent of the performance observed on the training set. We thus let
$$
c^1_\text{train}(\bm{P}_{\text{train}}(\bm{\beta},\bm{\gamma}))\coloneqq a
$$
where the best-fit is calculated as explained in \cref{sec:appCalcBestFit_expFit}. This choice of this cost function is better justified because in the extrapolation to higher dimensions we are mostly interested in its asymptotical performance, i.e. the better exponent in the extrapolation, the more suitable is Fixed-angle (CM-)QAOA for solving SVP instances and this is reflected in the cost function used in the training. The disadvantage is that it may happen that an optimiser finds angles $\bm{\beta},\bm{\gamma}$ so that the results on the training set $\bm{P}_\text{train}(\bm{\beta},\bm{\gamma})=\left\{P_\text{train}^{m_{\text{start}}}(\bm{\beta},\bm{\gamma}),P_\text{train}^{m_{\text{start}+1}}(\bm{\beta},\bm{\gamma}),\dots,P_\text{train}^{m_{\text{end}}}(\bm{\beta},\bm{\gamma})\right\}$, where $P_\text{train}^{m_{\text{i}}}(\bm{\beta},\bm{\gamma})$ is a pair
$$P_\text{train}^{m_{\text{i}}}(\bm{\beta},\bm{\gamma})=\left(\begin{array}{c}
	\text{averaged output of Fixed-angle (CM-)QAOA over }\\
	\bm{\mathcal{L}}^{m_i}_{\text{train}}\text{ with angles }\bm{\beta},\bm{\gamma}
\end{array}, m_i\right)$$,
do not follow an exponential function while the best-fit of $a,b$ to the function $2^{-am+b}$ returns low $a$. Such angles were observed not to extrapolate well. However, this behaviour has been observed in around a third of the training procedures. They showed a non-efficient performance with too little overlap over the solution, therefore they were discarded. In the remaining ones, the found angles $\bm{\beta},\bm{\gamma}$ not only found a small $a$, but also provided results that roughly estimated an exponential function and the extrapolation confirmed the trend.

To avoid over-fitting, we used a modified cost function \(c^\text{1,rand}_\text{train}(\cdot)\) where each iteration of the optimisation has access to only a random subset of both the training dimensions and also to only a random subset of the instances of these dimensions. We experimentally observed that a range about 80\%-90\% for both of the cases performed the best.

\subsubsection{\(c^2_\text{train}\): Maximising the squared difference from the random guess}

Another choice for  $c_\text{train}(\cdot)$ that we observed to be working particularly well works by maximising normalised squared distance of the training results $\bm{P}_\text{train}(\bm{\beta},\bm{\gamma})$ from the function $2^{-m}$ which represents the time complexity of solving $SVP$ by a naive brute-force enumeration attack.
\begin{align*}
	c^2_\text{train}(\bm{P}_{\text{train}}(\bm{\beta},\bm{\gamma})):=&\sum_{m_\text{start}\leq m\leq m_\text{end}}\big[2^m\big(P_\text{train}^{m_{\text{i}}}(\bm{\beta},\bm{\gamma})_{[1]}-2^{-m}\big)\big]^2\\
	&\ \ \ \ \text{where the $2^m$ coefficient serves as a normalisation}\\
	&\ \ \ \ \text{to give the difference in each dimension the equal weight}\\
	=&\sum_{m_\text{start}\leq m\leq m_\text{end}}\big[2^mP_\text{train}^{m_{\text{i}}}(\bm{\beta},\bm{\gamma})_{[1]}-1\big]^2
\end{align*}
The notation $P_\text{train}^{m_{\text{i}}}(\bm{\beta},\bm{\gamma})_{[1]}$ denotes the first element of the pair $P_\text{train}^{m_{\text{i}}}(\bm{\beta},\bm{\gamma})$ which is the averaged output of Fixed-angle (CM-)QAOAover $\bm{\mathcal{L}}^{m_i}_{\text{train}}\text{ with angles }\bm{\beta},\bm{\gamma}$.

The performance of $c^2_\text{train}(\cdot)$ was observed to be comparable to $c^1_\text{train}(\cdot)$ in terms of the asymptotic speedup observed in the training dataset. Its advantage is that it is less prone to peculiar results that provide an interesting best-fit asymptote but do not follow an exponential function. Its drawback is that it is less justified compared to \(c^1_\text{train}(\cdot)\) because it is only concerned about an improvement (that can be a constant improvement) over time complexity rather than an asymptotical scaling.

Similarly as with \(c^\text{1,rand}_\text{train}(\cdot)\), we have also found it beneficial to avoid over-fitting by using a randomised version \(c^\text{2,rand}_\text{train}(\cdot)\) with the identical randomisation procedure as with \(c^\text{1,rand}_\text{train}(\cdot)\), i.e. each iteration of the optimisation has access to only a random subset of both the training dimensions and also to only a random subset of the instances of these dimensions. Again, we experimentally observed that a range about 80\%-90\% for both of the cases performed the best.

\subsection{Exponential fitting for extrapolation of Fixed-angle (CM-)QAOA performance}\label{sec:appCalcBestFit_expFit}
Given $\bm{P}(\bm{\beta},\bm{\gamma})=\left\{P^{m_0}(\bm{\beta},\bm{\gamma}),P^{m_1}(\bm{\beta},\bm{\gamma}),\dots,P^{m_{n-1}}(\bm{\beta},\bm{\gamma})\right\}$ where $$P^{m_{\text{i}}}(\bm{\beta},\bm{\gamma})=\left(\begin{array}{c}
		\text{averaged output of Fixed-angle (CM-)QAOAover }\\
		\bm{\mathcal{L}}^{m_i}\text{ with angles }\bm{\beta},\bm{\gamma}
		\end{array}, m_i\right)$$
we use analytical approach to find the best-fit of $P^{m_{\text{i}}}(\bm{\beta},\bm{\gamma})_{[1]}$ to the function $$f(P^{m_{\text{i}}}(\bm{\beta},\bm{\gamma})_{[2]})=2^{-aP^{m_{\text{i}}}(\bm{\beta},\bm{\gamma})_{[2]}+b}$$ by determining the constants $a,b$. The notation $P^{m_{\text{i}}}(\bm{\beta},\bm{\gamma})_{[k]}$ denotes the $k$-th element of the pair $P^{m_{\text{i}}}(\bm{\beta},\bm{\gamma})$.

We use a widely-used analytical approach to exponential fitting with the least-squares method \cite{leastSquaresFit}. Given a function $f(m)=2^{-am+b}$ and the pairs $(y_i, m_i)$ to be fitted we proceed to find an analytical solution to constants $a,b$ that by taking the logarithms on both sides minimise the linearised version of problem:
\begin{equation}
(a,b)=\argmin_{a,b}\sum_{i=0}^{n-1}\bigl( y_i- f(m_i) \bigr)^2\approx\argmin_{a,b}\sum_{i=0}^{n-1}\bigl( \log_2 y_i- \log_2 f(m_i) \bigr)^2=\argmin_{a,b}\sum_{i=0}^{n-1}\bigl( \log_2 y_i+am_i-b \bigr)^2
\end{equation}
To find the minimum we calculate the first partial derivatives with respect to $a$ and $b$ and set them to $0$:
\begin{align}
	0&=\frac{\partial}{\partial b}\sum_{i=0}^{n-1}\bigl( \log_2 y_i+am_i-b \bigr)^2=-2\sum_{i=0}^{n-1}\bigl( \log_2 y_i+am_i-b \bigr)\nonumber\\
	&\implies \sum_{i=0}^{n-1} \log_2 y_i+\sum_{i=0}^{n-1}am_i-\sum_{i=0}^{n-1}b =0\nonumber\\
	&\implies nb=\sum_{i=0}^{n-1} \log_2 y_i+\sum_{i=0}^{n-1}am_i\nonumber\\
	&\implies b=\frac{\sum_{i=0}^{n-1} \log_2 y_i+\sum_{i=0}^{n-1}am_i}{n}\label{eq:b_partial}
\end{align}
\begin{align}
	0&=\frac{\partial}{\partial a}\sum_{i=0}^{n-1}\bigl( \log_2 y_i+am_i-b \bigr)^2=2\sum_{i=0}^{n-1}\bigl( \log_2 y_i+am_i-b \bigr)m_i\nonumber\\
	&\implies \sum_{i=0}^{n-1} \log_2 y_i m_i+\sum_{i=0}^{n-1}am_i^2-b\sum_{i=0}^{n-1}m_i =0\nonumber\\
	&\implies \sum_{i=0}^{n-1} \log_2 y_i m_i+\sum_{i=0}^{n-1}am_i^2-\frac{\sum_{i=0}^{n-1} \log_2 y_i+\sum_{i=0}^{n-1}am_i}{n}\sum_{i=0}^{n-1} m_i =0\nonumber\\
	&\implies n\sum_{i=0}^{n-1} \log_2 y_i m_i+na\sum_{i=0}^{n-1}m_i^2-\sum_{i=0}^{n-1} \log_2 y_i\sum_{i=0}^{n-1} m_i-a\left[\sum_{i=0}^{n-1}m_i\right]^2=0\nonumber\\
	&\implies na\sum_{i=0}^{n-1}m_i^2-a\left[\sum_{i=0}^{n-1}m_i\right]^2=\sum_{i=0}^{n-1} \log_2 y_i\sum_{i=0}^{n-1} m_i-n\sum_{i=0}^{n-1} \log_2 y_i m_i\nonumber\\
	&\implies a=\frac{\sum_{i=0}^{n-1} \log_2 y_i\sum_{i=0}^{n-1} m_i-n\sum_{i=0}^{n-1} \log_2 y_i m_i}{n\sum_{i=0}^{n-1}m_i^2-\left[\sum_{i=0}^{n-1}m_i\right]^2}\label{eq:calculated_a}
\end{align}
\begin{align}
	&\ \ \ \text{substituting \Cref{eq:calculated_a} into \Cref{eq:b_partial}}\nonumber\\
	&\implies b=\frac{\sum_{i=0}^{n-1} \log_2 y_i+\frac{\sum_{i=0}^{n-1} \log_2 y_i\sum_{i=0}^{n-1} m_i-n\sum_{i=0}^{n-1} \log_2 y_i m_i}{n\sum_{i=0}^{n-1}m_i^2-\left[\sum_{i=0}^{n-1}m_i\right]^2}\sum_{i=0}^{n-1}m_i}{n}\nonumber\\
	&\implies b=\frac{\sum_{i=0}^{n-1} \log_2 y_i}{n}+\frac{\sum_{i=0}^{n-1} \log_2 y_i\left[\sum_{i=0}^{n-1} m_i\right]^2-n\sum_{i=0}^{n-1} \log_2 y_i m_i\sum_{i=0}^{n-1} m_i}{n^2\sum_{i=0}^{n-1}m_i^2-n\left[\sum_{i=0}^{n-1}m_i\right]^2}\nonumber\\
	&\implies b=\frac{\sum_{i=0}^{n-1} \log_2 y_i\left(n\sum_{i=0}^{n-1}m_i^2-\left[\sum_{i=0}^{n-1}m_i\right]^2\right)+\sum_{i=0}^{n-1} \log_2 y_i\left[\sum_{i=0}^{n-1} m_i\right]^2}{n^2\sum_{i=0}^{n-1}m_i^2-n\left[\sum_{i=0}^{n-1}m_i\right]^2}\nonumber\\
	&\text{\ \ \ \ \ \ \ \ \ \ \ \ \ \ \ \ \ \ }-\frac{n\sum_{i=0}^{n-1} \log_2 y_i m_i\sum_{i=0}^{n-1} m_i}{n^2\sum_{i=0}^{n-1}m_i^2-n\left[\sum_{i=0}^{n-1}m_i\right]^2}\nonumber\\
	&\implies b=\frac{n\sum_{i=0}^{n-1} \log_2 y_i\sum_{i=0}^{n-1}m_i^2-n\sum_{i=0}^{n-1} \log_2 y_i m_i\sum_{i=0}^{n-1} m_i}{n^2\sum_{i=0}^{n-1}m_i^2-n\left[\sum_{i=0}^{n-1}m_i\right]^2}\nonumber\\
	&\implies b=\frac{\sum_{i=0}^{n-1} \log_2 y_i\sum_{i=0}^{n-1}m_i^2-\sum_{i=0}^{n-1} \log_2 y_i m_i\sum_{i=0}^{n-1} m_i}{n\sum_{i=0}^{n-1}m_i^2-\left[\sum_{i=0}^{n-1}m_i\right]^2}\label{eq:calculated_b}
\end{align}
We have thus found analytic expressions for \(a\) and \(b\) in \Cref{eq:calculated_a} and \Cref{eq:calculated_b} respectively\footnote{Note that this is a widely used approach for an exponential fit. Although the procedure has been widely expanded and accommodated to notation consistent with rest of the work, it cannot be claimed to be our contribution.}. Note that this method does not calculate the optimal \(a\) and \(b\) of least-squares exponential fitting. Because a logarithm was taken to linearise the exponential function, the calculated fit performs a least-squares fit where the errors are scaled logarithmically. This is in our case beneficial as it gives more weight to small errors for large \(m\) than to the large errors of small \(m\), i.e. the fit is more considerate about the fitting the tail of the function.

\bibliographystyle{alpha}
\bibliography{bibfile}

\end{document}